\pdfoutput=1
\documentclass[conference]{IEEEtran}
\IEEEoverridecommandlockouts
\usepackage{cite}
\usepackage{amsmath,amssymb,amsfonts}
\usepackage{algorithmic}
\usepackage{graphicx}
\usepackage{textcomp}
\usepackage{xcolor}
\usepackage{colortbl}
\usepackage{booktabs}
\usepackage{tabularx}
\usepackage{makecell}
\usepackage{multirow}
\usepackage{multicol}
\usepackage{xurl}
\usepackage{subcaption}
\usepackage {hyperref}
\hypersetup{hidelinks}
\usepackage{amsmath}
\usepackage{algorithm}

\begin{document}

\title{KAPG: Adaptive Password Guessing via Knowledge-Augmented Generation}

\author{
	\IEEEauthorblockN{
		Xudong Yang\IEEEauthorrefmark{1},
		Jincheng Li\IEEEauthorrefmark{1},
		Kaiwen Xing\IEEEauthorrefmark{1},
        Zhenjia Xiao\IEEEauthorrefmark{1},
        Mingjian Duan\IEEEauthorrefmark{2},
        Weili Han\IEEEauthorrefmark{2}
        and Hu Xiong\IEEEauthorrefmark{1}}
	\IEEEauthorblockA{\IEEEauthorrefmark{1}University of Electronic Science and Technology of China, Chengdu, China\\ Email: swithunyang918@gmail.com}
	\IEEEauthorblockA{\IEEEauthorrefmark{2}Fudan University, Shanghai, China}
}

\maketitle

\begin{abstract}
As the primary mechanism of digital authentication, user-created passwords exhibit common patterns and regularities that can be learned from leaked datasets. Password choices are profoundly shaped by external factors, including social contexts, cultural trends, and popular vocabulary. Prevailing password guessing models primarily emphasize patterns derived from leaked passwords, while neglecting these external influences---a limitation that hampers their adaptability to emerging password trends and erodes their effectiveness over time.

To address these challenges, we propose KAPG, a knowledge-augmented password guessing framework that adaptively integrates external lexical knowledge into the guessing process. KAPG couples internal statistical knowledge learned from leaked passwords with external information that reflects real-world trends. By using password prefixes as anchors for knowledge lookup, it dynamically injects relevant external cues during generation while preserving the structural regularities of authentic passwords. Experiments on twelve leaked datasets show that KnowGuess achieves average improvements of 36.5\% and 74.7\% over state-of-the-art models in intra-site and cross-site scenarios, respectively. Further analyses of password overlap and model efficiency highlight its robustness and computational efficiency. 
To counter these attacks, we further develop KAPSM, a trend-aware and site-specific password strength meter. Experiments demonstrate that KAPSM significantly outperforms existing tools in accuracy across diverse evaluation settings.
\end{abstract}



\section{Introduction}
\label{1}

Passwords remain the cornerstone of digital authentication and the most widely used barrier against unauthorized access to online services~\cite{bonneau2015passwords, wang2019birthday}. However, the ubiquity of passwords also makes them a prime target for attackers, who continue to develop sophisticated password guessing models to attempt to compromise user accounts, gain illicit access, and steal digital assets.

Existing password guessing methods—ranging from Markov~\cite{narayanan2005fast} and Probabilistic Context-Free Grammars (PCFGs)~\cite{weir2009password} to newer neural models—share a common foundation: They are primarily trained on leaked password datasets. By mining statistical distributions, character transformations, or structural templates from password datasets, these models aim to approximate how users construct passwords and generate guesses based on these patterns. Users often don't randomly combine characters when creating passwords; instead, they are heavily influenced by external knowledge. Popular culture, social events, and well-known platform names frequently infiltrate user choices: For example, \texttt{911} (derived from historical events), \texttt{covid19} or \texttt{ppe} (reflecting pandemic-related terms), and service names such as \texttt{qq} or \texttt{facebook} are all likely common in real-world passwords. Our empirical analysis of leaked datasets shows that a significant proportion of real-world passwords contain these external elements, either as complete words or as substrings of their variants. Existing password guessing models are often unable/difficult to guess passwords that contain external knowledge because this knowledge rarely appears in the password training set. For example, traditional models may require millions of guesses to crack password like \texttt{covid2019}, while human attackers with common sense in real-world contexts often quickly recognize its feasibility. This reveals the root of the problem: Models driven solely by static leaked data lack the ability to perceive and leverage external knowledge, resulting in a systematic omission of a large class of passwords that are strongly semantically related to real-world semantics.

One intuitive remedy is to directly incorporate external knowledge into the training set to expand coverage. For example, Weir et al.~\cite{weir2009password} used multiple external dictionaries in their PCFG model to expand the vocabulary coverage of generated passwords. These external dictionaries were used to replace letter variables in grammar rules during the generation phase to form the final candidate passwords. However, this introduces new biases and degradation: External knowledge often possesses semantic complexity and structural heterogeneity, significantly deviating from the statistical patterns of the passwords. Simply incorporating it dilutes the model's learning of the structure and distribution of real passwords, thereby reducing overall guessing accuracy. More importantly, this one-time injection fails to selectively utilize external information during generation, preventing precise selection based on specific candidates or context. On the one hand, leaked password data provides rich statistical and structural features; on the other hand, external knowledge provides semantic clues with broader coverage and close correlation with real-world trends. This leads to the core challenge of solving this problem: How to design a mechanism that effectively couples external knowledge with internal password knowledge, thereby preserving the advantages of internal statistical regularities while fully leveraging the complementary role of external semantic clues.

At a high level, password guessing can be formulated as a knowledge-augmented generation task: Internal knowledge encodes the structural regularities of real passwords, while external knowledge provides trend-aware lexical evidence that can be selectively incorporated during the guessing process. Formally, let the password space be $S \in cs^*$, where $cs$ is the set of possible characters (e.g., letters, digits and special characters). Internal knowledge comes from a corpus of leaked passwords. Define the mapping $\phi_{\text{int}}$, which represents the user's password creation behavior learned by traditional guessing models. Given an external knowledge base $Z$ and a knowledge query strategy $Q$, the retrieval operator $R$ generates the context related to the generation of candidates: $C(s)=R(Q(s),Z)$. Through a fusion function, we can obtain the coupling result of internal knowledge and external knowledge: \(F(\phi_{\text{int}}(s),\; \psi(s,C(s)),\; \xi(s,C(s)),\) where $\psi$ measures the alignment between the candidate and retrieved context, $\xi$ acts as an adaptive gate controlling the influence of external information. The induced guessing policy is then: \(\pi(s) \propto \exp(F(\phi_{\text{int}}(s),\; \psi(s,C(s)),\; \xi(s,C(s)))).\)

We acknowledge that many potential complexities are naturally avoided in this characterization. Extracting the internal knowledge of a password can be handled by any existing password guessing model (e.g., Markov or PCFG), as these models already provide strong coverage of password characteristics. Similarly, the space for query construction is limited: Retrieval can be based on prefixes, full strings, or structural templates. A key consideration is the compatibility between password internal knowledge extraction and query mechanisms. For instance, if the internal knowledge of a password is modeled at the structural level (e.g., via PCFG templates like L3D2S1), while queries are issued at the character level, the retrieved external knowledge may be inconsistent with the internal representation, thus undermining effective fusion. To mitigate this, our framework uses a shared embedding space (e.g., via one-hot vectors) for both internal and external representations. Among these options, we select prefix-based retrieval as it aligns with the user's gradual password construction process—users build passwords incrementally, influenced by semantic cues like trends—and enables incremental retrieval to reflect user behavior and dynamic adjustments during generation. This approach ensures the prefix acts as a query anchor to retrieve relevant full strings (or patterns), which then inform the probability of subsequent characters, rather than directly equating the prefix's standalone probability to the full strings. The alignment function $\psi$ simply measures the degree of match between a candidate password and the retrieved external knowledge—for example, testing whether the prefix \texttt{cov} is closely related to \texttt{covid-19}. The gating function $\xi$ acts as a regulator: It allows a stronger influence when the retrieved knowledge is highly relevant. Finally, the fusion function combines the internal predictions modeled from leaky passwords with external cues to produce a single distribution that guides the guessing process.

Based on this mechanism, we instantiated KAPG, a knowledge-augmented password guessing framework. The internal feature mapping $\phi_{\text{int}}$ is implemented by a baseline password Generator that learns statistical patterns from leaked password data; the retrieval operator $R$ and alignment $\psi$ are implemented as Retriever that provide relevant external knowledge; the gate $\xi$ and fusion mapping $F$ are jointly embodied in the Fusionizer module, which adaptively integrates the two sources during the guessing process. Specifically, the Generator learns features from the leaked password dataset and generates a baseline probability distribution for password guesses. The generator is a fourth-order Markov model that generates guessed passwords by learning transition probabilities between password prefixes and characters from the password data. Notably, it employs a back-off mechanism to address the sparse context problem, ensuring benchmark robustness. The Retriever encodes prefixes as feature vectors, significantly enhancing the expressive power of prefix queries and enabling the model to retrieve semantically related and structurally similar terms (e.g., capturing the similarity between \texttt{pass} and \texttt{p@ss}). Then, the 10 most similar terms are retrieved using Approximate Nearest Neighbour (ANN) search in Faiss.\footnote{Faiss (Facebook AI Similarity Search) is a library for efficient similarity search and dense vector clustering. Code repository: \url{https://github.com/facebookresearch/faiss}} The index uses L2 distance as the metric. This mechanism effectively identifies and supplements missing potential patterns in static training data by calculating the L2 distance between prefix vectors and knowledge item vectors. The Fusionizer module adaptively weights retrieval results based on the L2 distance scores obtained from ANN, fusing the generator's baseline probability distribution with external patterns retrieved by the Retriever to calibrate the password guessing distribution.

The guessing performance of KAPG is evaluated in multi-scenario experiments. It exhibits near-optimal cracking capabilities compared to existing state-of-the-art models. The analysis of password overlap rate and model efficiency further highlights its robustness. Notably, we further extend KAPG to naturally support dynamic password guessing (see Appendix~\ref{Avalanche cracking}), a capability proposed by Pasquini~\cite{pasquini2021improving} that aims to leverage successfully cracked passwords to dynamically adjust subsequent guesses. As KAPG has demonstrated its powerful capabilities in password guessing, it also poses greater challenges to attack defence (i.e., password strength meter~\cite{pasquini2020interpretable, xu2021chunk}). We have observed that most of the widely used password strength meter (PSM) still use static evaluation mechanisms, lack the ability to perceive popular trends, and find it difficult to adjust evaluation strategies in a timely manner based on changes in user behaviour. Pasquini et al. pointed out that a robust password model needs to adapt to the password distribution characteristics of its target user group~\cite{pasquini2024universal}, which is also applicable to PSM: \textbf{Only a strength meter that can adapt to the habits of users in its deployment environment can achieve more accurate and reliable security evaluations}. To this end, we have developed and implemented a PSM called KAPSM, which can dynamically adapt to the password distribution of the deployed website and realise trend perception and response to user behaviour evolution. Experiments have proven that KAPSM significantly improves assessment accuracy in multiple password datasets and simulated deployment environment, as well as having good customisation capabilities.

We summarize main contributions as follows.

\begin{itemize}

\item[$\bullet$] \textbf{Analysis of external knowledge-driven password construction}. A systematic analysis of multiple leaked password datasets is undertaken to reveal the influence of external knowledge in user-generated passwords. Specifically, we analyze external knowledge in passwords in terms of popular vocabulary, cultural context, and linguistic features. We demonstrate the critical role of external factors in shaping user password creation. These findings not only provide empirical evidence for understanding the multiple origins of user-created passwords, but also lay the foundation for designing more adaptive guessing models.

\item[$\bullet$] \textbf{A knowledge-augmented password guessing framework}. We propose KAPG, the first knowledge-augmented password guessing model. KAPG dynamically incorporates external knowledge—such as popular culture, trending vocabulary, and cultural contexts—into the guessing process, effectively bridging the gap between static password datasets and real-world factors that influence password creation. 
Experimental results show that KAPG achieves optimal cracking performance comparable to FLA and RFGuess while outperforming OMEN, PCFG, and CKL\_PCFG by 36.5$\%$ and 74.7$\%$ in intra-site and cross-site scenarios, respectively. Furthermore, KAPG achieves superior resource efficiency with balanced model size (1,091.2 MB), training time (64.3 s), and guessing speed (588,144.2 pwd/s). Moreover,  we explore dynamic password guessing based on KAPG in the appendix~\ref{Avalanche cracking}, which can further improve the password guessing cracking rate by 22.3$\%$.

\item[$\bullet$] \textbf{A security practice oriented to defence}. We designed and implemented KAPSM, an adaptive password strength evaluation framework. KAPSM can achieve trend awareness and website-specific strength assessment. Compared with mainstream PSMs, experimental results on multiple password datasets show that KAPSM's password strength assessment accuracy has improved by up to 51.4$\%$. Furthermore, in our simulated deployment environment, KAPSM is able to achieve continuous assessment optimisation by gradually integrating user passwords from the target website, resulting in an increase in password strength assessment accuracy of 9.5$\%$.

\end{itemize}


\section{Preliminary}
\label{2}
\subsection{Notation}
Here, we summarize and clarify the key symbols in this paper, as detailed in Table~\ref{tab:notation}


\begin{table}[ht]
\centering
\setlength\tabcolsep{6pt}
\caption{Notations Summary.}
\label{tab:notation}
\resizebox{\columnwidth}{!}{
\begin{tabular}{cc}
\toprule
Notation & Description \\
\midrule
\thead{\( pwd \)} & \thead{Password} \\
\thead{\( c_i \)} & \thead{The \( i \)-th character in a password} \\
\thead{\( p_{\text{int}} \)} & \thead{The internal probability distribution computed \\ by the Generator} \\
\thead{\( T \)} & \thead{The transition matrix of the Generator} \\
\thead{\( Z \)} & \thead{The external knowledge base, a collection \\ of popular terms or patterns} \\
\thead{\( z \)} & \thead{An individual knowledge item (e.g., a term \\ or pattern) in the external knowledge base \( Z \)} \\
\thead{\( f \)} & \thead{The embedding function converting character \\ sequences into vector representations} \\
\thead{\( q \)} & \thead{The query vector, generated from an input sequence \\ via the embedding function for retrieval} \\
\thead{\( Z_i \)} & \thead{The collection of knowledge items retrieved from \\ the external knowledge base \( Z \) most relevant \\ to the prefix \( c_1, \ldots, c_{i-1} \), used to \\ predict the \( i \)-th character} \\
\thead{\( p_{\text{ext}} \)} & \thead{The external probability distribution derived from \\ retrieved patterns to enhance prediction} \\
\thead{\( cs \)} & \thead{The character set, representing the collection \\ of all possible characters (i.e., vocabulary)} \\
\bottomrule
\end{tabular}}
\end{table}

\subsection{Password guess}
Here, we review the development of data-driven password guessing techniques, highlighting their evolution and improvements while discussing their inherent adaptability challenges.

One of the earliest data-driven approaches to password guessing is the Markov chain-based probabilistic model. In 2005, Narayanan et al.~\cite{narayanan2005fast} proposed Markov guessing model that captures inter-character dependencies to predict the probability of password generation. The core idea is to use preceding characters to estimate the likelihood of the next character. For a password \texttt{123abc@}, its generation probability (using 3-order as example) can be expressed as the product of conditional probabilities: 
$P(\texttt{123abc@}) = P(\texttt{1}|\texttt{\#\#\#})\times \ldots\times P(\texttt{@}|\texttt{abc})$,
where \texttt{\#} denotes the start symbol of the Markov model. Higher-order models capture richer contextual information, improving guessing accuracy, but they are prone to overfitting. To address this, subsequent research introduced smoothing and normalization techniques. For instance, Ma et al.~\cite{ma2014study} employed Laplace smoothing and normalization to mitigate overfitting, enhancing the model’s applicability in sparse data scenarios. Further progress includes generating probability descending guessing password~\cite{durmuth2015omen} and expanding prefix context~\cite{xie2022wordmarkov}, which have bolstered the guessing power of Markov models. 
Concurrently, PCFG offer an alternative data-driven approach to modeling password patterns. In 2009, Weir et al.~\cite{weir2009password} introduced a PCFG-based password guessing algorithm that parses passwords into structural templates and terminals. The structural template consists of basic components (letter segments L, digit segments D, and special character segments S), while terminals represent the specific characters. The probability of a password is calculated as the product of the structural probability and the terminal probabilities (taking password \texttt{123abc@} as an example): 
$P(\texttt{123abc@}) = P(L_3D_3S_1) \times P(L_3 \rightarrow \texttt{123}) \times P(D_3 \rightarrow \texttt{abc}) \times P(S_1 \rightarrow \texttt{@})$. 
PCFG effectively captures structural patterns in passwords, but early models struggled with complex patterns. Subsequent research introduced several enhancements, such as keyboard pattern modeling~\cite{houshmand2015next}, natural language instantiation~\cite{xu2021chunk}, and semantic analysis to uncover deeper structural features~\cite{veras2014semantic}. 
In recent years, the advent of deep learning has brought significant advancements to password guessing while still operating within a data-driven paradigm. In 2016, Melicher et al.\cite{melicher2016fast} pioneered the use of Long Short-Term Memory (LSTM) networks for password guessing, leveraging their sequence modeling capabilities to significantly improve cracking rates compared to traditional PCFG and Markov models. Building on advances in deep learning, Hitaj et al.\cite{hitaj2019passgan} introduced PassGAN in 2019, a Generative Adversarial Network (GAN) that learns the distribution of real passwords to generate high-quality guesses closely matching the statistical characteristics of real-world password datasets. Pasquini et al.~\cite{pasquini2021improving} further refined this approach by adopting Wasserstein GAN (WGAN) and Context Wasserstein Autoencoder (CWAE), mitigating mode collapse issues in GAN training and enhancing the diversity and authenticity of generated passwords. Additionally, Variational Autoencoders (VAEs) have been employed to learn latent distributions of passwords for more realistic generation~\cite{yang2022vaepass}. However, mapping from continuous to discrete spaces often leads to precision loss. To address this, recent efforts have shifted toward language model-based approaches, such as PassGPT~\cite{rando2023passgpt}, which utilizes Generative Pre-trained Transformers (GPT), and PassBERT~\cite{xu2023improving}, based on bidirectional encoder representations, leveraging large-scale pre-training and contextual awareness to further improve the accuracy and efficiency of password guessing.

Despite significant advancements in cracking performance, structural modeling, and generation diversity achieved by data-driven password guessing methods, their inherent limitations hinder practical applicability. Traditional models rely solely on static leaked password datasets, capturing only password's structural and statistical features. However, user-generated passwords are influenced by public information, including cultural elements and external patterns. These models lack mechanisms to incorporate such external knowledge, resulting in limited coverage and suboptimal guessing accuracy. Moreover, these models require retraining to adapt to external information or updated datasets from recent breaches. The resulting resource requirements often discourage practical password guessing, especially in real-time scenarios where efficiency is critical.

\section{Analysis of password associations with external knowledge}
\label{3}

\subsection{Datasets}

\begin{table}[htbp]\centering
\setlength\tabcolsep{0.8pt}
    \caption{Summary of the password datasets used in this
paper.}
    \label{Datasets}
    \resizebox{\columnwidth}{!}{
    \begin{tabular}{ccccc}
        \toprule
        Datasets & Web service & Language & Leaked time & Total passwords \\
        \midrule
        Rockyou~\cite{Rockyou} & Social forum & English & Dec. 2009 & 32,581,870 \\
        CSDN~\cite{csdn} & Programmer forum & Chinese & Dec. 2011 & 6,428,277 \\
        Dodonew~\cite{Dodonew} & E-commerce & Chinese & Dec. 2011 & 16,258,891 \\

        7K7K~\cite{7K7K} & Mini-games & Chinese & Dec. 2011 & 8,160,101 \\
        
        Tianya~\cite{Tianya} & Social forum & Chinese & Dec. 2011 & 30,179,474 \\

        Linkedin~\cite{LinkedIn} & Job hunting & English & Jan. 2012 & 54,656,615\\
        Gmail~\cite{Gmail} & Email & English & Sep. 2014 & 1,421,372 \\
        Netease~\cite{Netease} & Email & Chinese & Dec. 2015 & 6,392,568 \\
        Taobao~\cite{Taobao} & E-commerce & Chinese & Feb. 2016 & 15,072,418\\
        Neopets~\cite{Neopets} & Pet information & English & Jul. 2016 & 67,672,205 \\
        Clixsense~\cite{Clixsense} & Paid task platform & English & Jun. 2016 & 2,222,045\\
        Mathway~\cite{mathway} & Learning support & English & May. 2020 & 16,457,736 \\
        \bottomrule
    \end{tabular}}
\end{table}

In this study, we choose twelve real password datasets for experiments (as shown in Table~\ref{Datasets}), including six Chinese datasets and six English datasets. These datasets present different characteristics due to the differences in their language environments and service types. All the data come from public sources on the Internet and are widely used in the field of password security research~\cite{xu2021chunk,wang2023pass2edit,xu2019coarse,xiu2024pointerguess}. Although these datasets involve users' sensitive credential information, we always adhere to strict ethical norms during the research process, and are committed to preventing privacy leakage and potential secondary damages caused by the use of the data.

In the preprocessing phase, we cleaned and filtered the raw password data to ensure that it was relevant to user-generated passwords.  Only passwords between 5 and 20 characters in length are retained, as this range captures common user-generated passwords that strike a balance between memorability and security. Passwords outside this range are typically too short (which may reflect non-user artifacts such as test data) or too long (which may be system-generated passwords) to represent user behavior. In addition, only characters in the ASCII code range of 32 to 126 (printable keyboard characters, including letters, digits, and special characters) are retained~\cite{wang2023password}.

\subsection{Analysis}
Here, we investigate the presence of external knowledge (e.g., popular vocabulary and cultural references) in user-generated passwords to validate the hypothesis that such information influences password construction. By analyzing leaked password datasets, we quantify the proportion of passwords containing external elements and characterize their semantic patterns. This analysis is motivated by the potential of external knowledge to provide clues for password cracking. 

We compiled a collection of popular Chinese and English vocabulary terms spanning from 2000 to 2020 (see Table~\ref{table:popular} for an example of popular words in 2018-2020). In breached password datasets, we calculated the proportion of these popular terms (dated prior to each dataset’s breach) to assess their impact on password evolution. Concretely, for English popular terms predating 2009, we identified 94 items, of which 48.9$\%$ appeared in the Rockyou dataset. The top-five most frequent terms were \texttt{im}, \texttt{911}, \texttt{gm}, \texttt{cool}, and \texttt{sweet}. These terms are often tied to the prevailing pop culture of the time (e.g., \texttt{cool} as a fashionable expression) or significant societal events (e.g., \texttt{911}), suggesting that users tend to incorporate familiar cultural or social elements into their passwords. 
For the popular Chinese terms before 2011, we recorded 120 terms and analysed their usage in CSDN, Dodonew, 7K7K and Tianya, the four 2011 leaked password datasets. In the CSDN dataset, these terms appeared in 39.2$\%$ of the cases, with the top five being \texttt{he}, \texttt{han}, \texttt{qq}, \texttt{yumen} and \texttt{xiaozi}. In the Dodonew dataset, 42.5$\%$ of the terms occur, led by \texttt{he}, \texttt{qq}, \texttt{911}, \texttt{lei} and \texttt{mba}. In the 7K7K dataset, 42.5$\%$ of the terms appeared, led by \texttt{he}, \texttt{han}, \texttt{911}, \texttt{qq} and \texttt{ding}. In the Tianya dataset, 45.0$\%$ of the terms occur, led by \texttt{he}, \texttt{han}, \texttt{911}, \texttt{qq} and \texttt{di}. Notably, \texttt{qq} was the name of a widely used social networking platform at the time, and the common prominence in the four datasets suggests that Chinese users were influenced by contemporary trends when crafting passwords. In addition, differences in lexical incorporation such as \texttt{yumen} and \texttt{xiaozi} in CSDN and \texttt{mba} and \texttt{911} in Dodonew are likely to reflect platform-specific user demographics. 
For English popular terms prior to 2012, we recorded 130 entries, of which 52.4$\%$ appeared in the Linkedin dataset, with the top-five being \texttt{911}, \texttt{sweet}, \texttt{matrix}, \texttt{dude} and \texttt{ maverick}. We recorded 150 English popular words for the period before 2014, of which 56.2$\%$ appeared in the Gmail dataset, and the top five were \texttt{gm}, \texttt{911}, \texttt{3d}, \texttt{cool}, and \texttt{sweet}. 
For Chinese popular terms predating 2015, we analyzed 160 items, of which 37.5$\%$ surfaced in the Netease dataset, with the top-five being \texttt{he}, \texttt{han}, \texttt{911}, \texttt{baobao}, and \texttt{nixi}. For Chinese and English popular words before 2016, we recorded 170 entries each, of which 41.8$\%$ appeared in the Taobao dataset, with the top five being \texttt{he}, \texttt{qq}, \texttt{911}, \texttt{baobao}, and \texttt{dav}. In Clixsense, 46.3$\%$ of the records appeared, and the top five were \texttt{im}, \texttt{911}, \texttt{gm}, \texttt{3d} and \texttt{luv}. In Neopets, 40.7$\%$ of the records appeared, with the top five being \texttt{im}, \texttt{cool}, \texttt{luv}, \texttt{911} and \texttt{dude}. 
Finally, for English popular terms predating 2020, we examined 201 items, with 46.3$\%$ appearing in the Mathway dataset. The top-five were \texttt{im}, \texttt{cool}, \texttt{ppe}, \texttt{lvu}, and \texttt{var}. Intriguingly, \texttt{ppe} (personal protective equipment) likely relates to the 2020 pandemic, underscoring the immediate influence of societal events on password selection.

\begin{table*}[htbp]\centering
\setlength\tabcolsep{5pt}
    \caption{Example of popular words in 2018, 2019 and 2020.}
    \label{table:popular}
    
    \begin{tabular}{cccc}
        \toprule
         & 2018(Chinese/English) & 2019(Chinese/English) & 2020(Chinese/English) \\
        \midrule
        Word-1 & \texttt{mingyungongtong}/\texttt{single-use} & \texttt{wodezuguo}/\texttt{quidproquo} & \texttt{tuopin}/\texttt{covid-19} \\
        Word-2 &  \texttt{jinli}/\texttt{backstop} & \texttt{jinseshinian}/\texttt{impeach} & \texttt{kouzhao}/\texttt{infodemic} \\
        Word-3 & \texttt{dianxiaoer}/\texttt{floss} & \texttt{xuexiqiangguo}/\texttt{crawdad} & \texttt{kangyi}/\texttt{self-isolation} \\
        Word-4 & \texttt{jiaokeshu}/\texttt{gammon} & \texttt{zhongmei}/\texttt{egregious} & \texttt{nixingzhe}/\texttt{self-quarantine} \\
        Word-5 & \texttt{guanxuan}/\texttt{gaslight} & \texttt{fendouzhe}/\texttt{clemency} & \texttt{jiankangma}/\texttt{social distancing} \\
        Word-6 & \texttt{querenyanshen}/\texttt{metoo} & \texttt{yinghe}/\texttt{snitty} & \texttt{xinguan}/\texttt{wfh} \\
        Word-7 & \texttt{tuiqun}/\texttt{plogging} & \texttt{lajifenlei}/\texttt{tergiversation} & \texttt{quntimianyi}/\texttt{ppe} \\
        Word-8 & \texttt{foxi}/\texttt{var} & \texttt{shifanqu}/\texttt{camp} & \texttt{rongduan}/\texttt{flatten thecurve} \\
        Word-9 & \texttt{juying}/\texttt{vegan} & \texttt{jianfu}/\texttt{exculpate} &  \texttt{meiguodaxuan}/\texttt{social recession} \\
        Word-10 & \texttt{gangjing}/\texttt{whitewash} & \texttt{tainanle}/\texttt{the} & \texttt{kebi}/\texttt{elbow bump} \\
        
        \bottomrule
    \end{tabular}
\end{table*}

Moreover, we investigated the impacts of cultural backgrounds and linguistic characteristics on the generation and evolution of passwords. To this end, we compiled a diverse set of representative entries, including high-frequency Pinyins/words in Chinese and English (e.g., \texttt{woaini} and \texttt{love}), popular English names (e.g., \texttt{alice}), Chinese surnames (e.g., \texttt{yang}), as well as frequently used job titles (e.g., \texttt{president}), book titles (e.g., \texttt{three-body}), game names (e.g., \texttt{dota}), and movie titles (e.g., \texttt{flipped}). These entries were derived from publicly available cultural and statistical data. 
By comparing twelve password datasets, we quantify the proportion of these external words used in password generation. The experimental results show that a large number of external words appear either as complete passwords or as components of passwords in different passwords. Specifically, the proportion of external words used as full passwords and password strings in Rockyou is 17.1$\%$ and 41.1$\%$, respectively. 
In the subsequent leaks of English-language sites Linkedin, Clixsense, Gmail and Neopets, 21.1$\%$, 16.6$\%$, 23.9$\%$ and 13.4$\%$ used external words as the full password, and 44.8$\%$, 42.7$\%$, 46.2$\%$ and 40.1$\%$ used external words as a substring of the password, respectively. 
In the recently leaked Mathway, this drops to 12.5$\%$ and 41.2$\%$. This reflects the preference of early users on the Internet for simple, easy-to-remember password patterns. This preference stems from a superficial understanding of security risks and user concerns about the ease, rather than the complexity, of password creation. 
In CSDN, Dodonew, 7K7K and Tianya, the percentage of using external words as full passwords is 2.3$\%$, 3.46$\%$, 5.7$\%$ and 5.8$\%$, while the percentage of using external words as substrings reaches 23.8$\%$, 21.37$\%$, 23.9$\%$ and 24.4$\%$. 
Similarly, in the Netease, these proportions were 4.7$\%$ and 22.7$\%$, respectively. In the 2016 Taobao leak, the percentage of external words used as full passwords and password strings was 4.7$\%$ and 28.9$\%$, respectively.

Based on our analysis, we find that user passwords exhibit certain patterns that partially overlap with those observed in leaked password datasets, while also reflecting influences from external knowledge sources. These empirical findings not only confirm the diversity of password generation patterns, but also highlight the limitations of traditional static password cracking methods in terms of adaptability and predictive power. Therefore, it has become imperative to improve the efficiency and accuracy of password guessing by integrating external knowledge.

\section{KAPG: a knowledge-augmented password guessing framework}
\label{4}

\subsection{Password modeling}
We now detail how password guessing is modeled within our proposed knowledge-augmented framework. 
Password generation can be naturally viewed as an incremental process, in which partial prefixes are progressively expanded until a complete candidate is produced. 
In our formulation, the generation of each character is jointly influenced by the statistical regularities learned from leaked passwords and the external lexical cues that reflect real-world trends and semantics.

Formally speaking, given a prefix $c_{i},c_{i+1},c_{i+2}$ (e.g., \texttt{cov}), the baseline password model $\phi_{\text{int}}$ generates a probability distribution for the next character. For example, $P(\texttt{a}|\texttt{cov})=0.35$, $P(\texttt{e}|\texttt{cov})=0.25$, and $P(\texttt{i}|\texttt{cov})=0.05$, which reflect only the characteristics observed in the password data. Simultaneously, the prefix $c_{i},c_{i+1},c_{i+2}$ is sent as a query $Q(c_{i},c_{i+1},c_{i+2})$ to an external knowledge base $Z$, generating a retrieval context $C(c_{i},c_{i+1},c_{i+2})=R(Q(c_{i},c_{i+1},c_{i+2}),Z)$. For example, the retriever might return entries such as \texttt{covid}, \texttt{covid19}, and \texttt{corona}. The alignment function $\psi(c_{i},c_{i+1},c_{i+2},C(c_{i},c_{i+1},c_{i+2}))$ evaluates the match between the prefix and each retrieved candidate word—for example, \texttt{covid} aligns strongly with \texttt{cov}, while \texttt{corona} aligns more weakly. Then, the gating function $\xi(c_{i},c_{i+1},c_{i+2},C(c_{i},c_{i+1},c_{i+2}))$ assigns weights based on the correlation, for example, the correlation between \texttt{covid} and the current prefix is 0.8, and the correlation between \texttt{corona} and the current prefix is 0.2. Finally, these weighted external signals are fused with the internal distribution $\phi_{\text{int}}$ via the mapping $F(\cdot)$. Specifically, the baseline distribution \{``a'':0.35,``e'':0.25,``i'':0.05,\dots\} is calibrated by external cues to \{``a'':0.20,``e'':0.10,``i'':0.35,\dots\}. Characters are sampled from the calibration distribution to generate the next character. This process is iterative: Each newly generated character is extended with a prefix (\texttt{cov} $\rightarrow$ \texttt{covi} $\rightarrow$ \texttt{covid}), and at each stage, the retrieval and fusion mechanism dynamically updates the probability distribution of the next character.

\subsection{The Architecture of KAPG}
\begin{figure}[!htbp]
    \centering
    \includegraphics[scale=0.24]{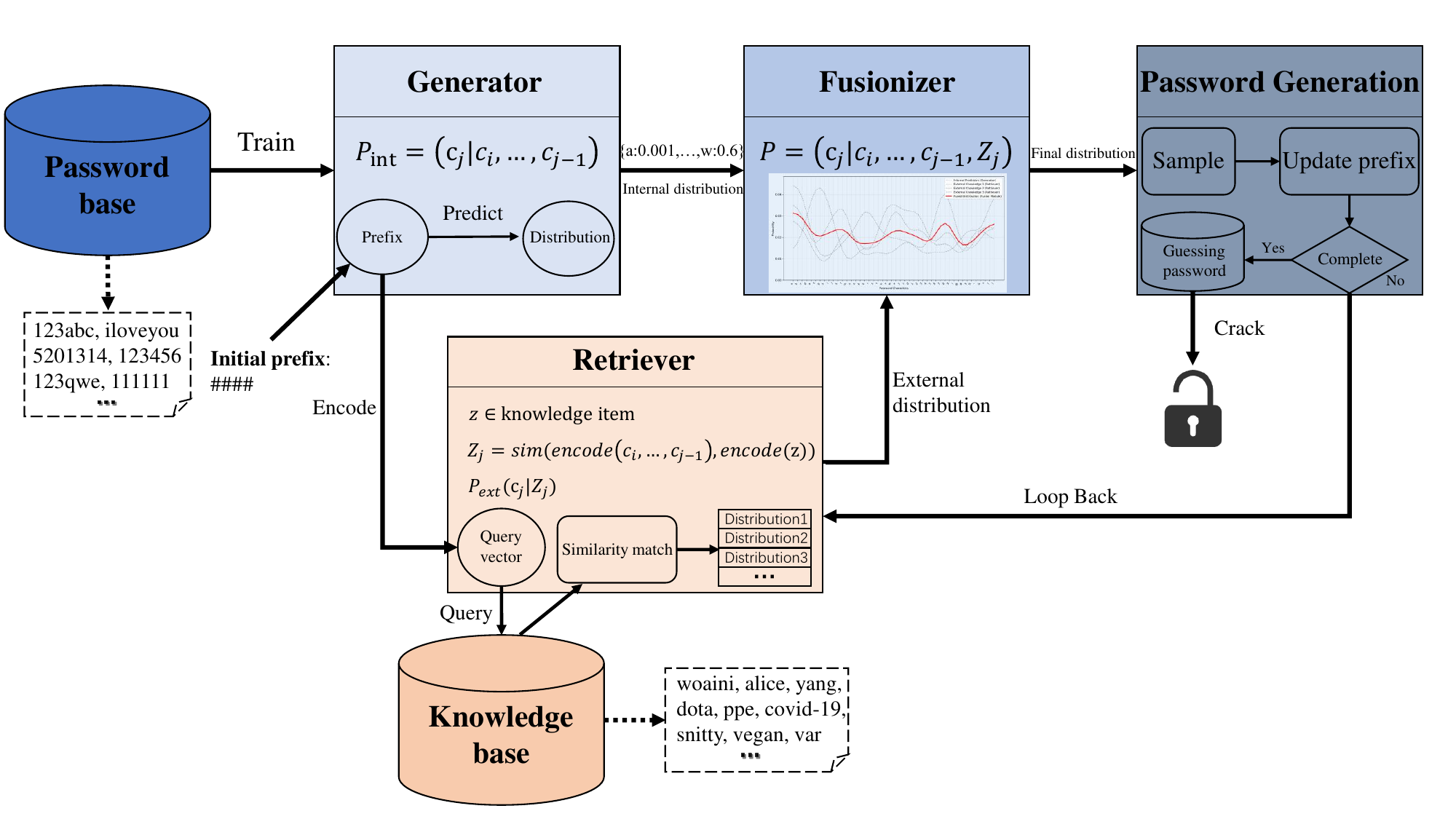}
    \caption{Workflow of KAPG for cracking passwords.}
    \label{fig:ragguess}
\end{figure}

This section elaborates on the system design of KAPG and the training process of each module, demonstrating how it achieves adaptive password guessing. KAPG comprises three core modules: Generator, Retriever, and Fusionizer (see Figure~\ref{fig:ragguess}).

\subsubsection{Generator module}

The Generator serves as the foundational component of KAPG, responsible for generating initial character probability distributions based on the current prefix. It employs a Markov-based sequential prediction model, leveraging local dependencies between characters extracted from static training data. The choice of Markov model over neural network alternatives (e.g., LSTM) is deliberate and theoretically grounded—despite LSTMs' impressive performance in recent password guessing research, our analysis reveals that Markov's explicit sparsity handling provides superior synergy (see Appendix~\ref{lstm_selection} for detailed theoretical and empirical justification).

Given a prefix $c_1, \ldots, c_{i-1}$, the Generator computes the internal probability distribution: $P_{\text{int}}(c_i | c_1, \ldots, c_{i-1})$, which encapsulates inherent patterns in user-generated passwords by analyzing local transition tendencies in the training data. Implemented as a 4th-order Markov chain, the Generator constructs a transition matrix $T$ from frequency statistics, where $T(c_i | c_{i-4}, c_{i-3}, c_{i-2}, c_{i-1})$ denotes the conditional transition probability given the prior four characters. 
The Generator constructs the transition table $T$ by directly counting the frequency of subsequent characters for prefixes in the static training dataset $D_{\text{train}} = \{pwd_1, pwd_2, \ldots, pwd_m\}$. For each prefix $c_{i-k}, \ldots, c_{i-1}$ (where $k$ ranges from 1 to 4), the transition probability is calculated as: $T(c_i | c_{i-k}, \ldots, c_{i-1}) = \frac{\text{count}(c_{i-k}, \ldots, c_{i-1}, c_i)}{\text{count}(c_{i-k}, \ldots, c_{i-1})}$, with probabilities for all prefixes precomputed and stored. To handle cases where high-order prefixes are absent from the training data, a backoff mechanism is employed: If a 4th-order prefix is missing, the model falls back to 3rd-, 2nd-, or 1st-order prefixes for probability estimation. 

\subsubsection{Retriever module}

The Retriever is responsible for enhancing the password guessing capability of KAPG by extracting contextually relevant patterns from an external knowledge base. Unlike the Generator that relies on static training data, the Retriever leverages dynamic external knowledge to capture real-time factors influencing user password creation.

The external knowledge base is built offline from an external knowledge dataset, which specifically collects external factors influencing password creation, such as popular vocabulary, cultural terms, and language patterns, complementing the static training dataset $D_{\text{train}}$. For each prefix, the system computes its subsequent character distribution, forming key-value pairs (i.e., \{\texttt{prefix: character distribution}\}). In practice, the prefix is extracted from the external dataset and transformed into embeddings via the one-hot encoder $f(\cdot)$, supporting a vocabulary of 94 characters ($|\text{vocab}| = 94$) with a maximum prefix length of 4 characters ($\text{max\_len} = 4$). The embedding matrix is defined as: $\text{embeddings} = [f(c) \text{ for } c \text{ in prefixes}]$, with a dimensionality of $d = \text{max\_len} \times |\text{vocab}| = 376$. Faiss employs the $\text{IndexFlatL2}$ index to store these embeddings, using L2 distance for efficient approximate nearest neighbor search, enabling rapid retrieval during inference.

The core idea of the retriever is to leverage password prefixes to retrieve highly relevant information from the knowledge base. We describe this process in two parts: How the password prefix is encoded for retrieval, and how the retrieved information is represented and utilized from the knowledge base. To efficiently capture the features of the prefix $c_{i-4}, \ldots, c_{i-1}$, the Retriever maps the prefix as a query vector $q = f(c_{i-4}, \ldots, c_{i-1})$ using a one-hot encoder to generate a high-dimensional embedding. This encoding represents each character sequence as a unique vector, preserving local dependencies. External knowledge is stored using the Faiss vector database with embedded representations of popular terms, cultural patterns, and language patterns. The retrieval process is formalized as: $Z_i = \arg\max_{z \in Z, |Z_i|=k} \text{sim}(q, f(z))$, where $z \in Z$ represents an individual knowledge item (e.g., a term or pattern) in the external knowledge base $Z$, $\text{sim}(\cdot, \cdot)$ is the L2 distance, and $k = 10$ specifies the number of relevant external knowledge items returned. The L2 distance was chosen as the similarity metric because it effectively measures the Euclidean difference between one-hot encoded vectors, ensuring that retrieved patterns are similar to the query prefix, which is crucial for predicting the next character in passwords influenced by user habits or trending terminology. The setting of $k = 10$ balances the trade-off between capturing sufficient contextual diversity and maintaining computational efficiency, based on our consideration that gains diminish after a small number of highly relevant searches.

\subsubsection{Fusionizer module}

The Fusionizer module integrates the Generator's internal predictions with the Retriever's external knowledge to produce the final character probability distribution. It computes the composite probability as: $P(c_i | c_{i-4}, \ldots, c_{i-1}, Z_i) = (1 - \lambda) P_{\text{int}}(c_i | c_{i-4}, \ldots, c_{i-1}) + \lambda P_{\text{ext}}(c_i | Z_i)$, where $P_{\text{ext}}(c_i | Z_i) = \sum_{z \in Z_i} w(z) P(c_i | z)$ is the external probability, with $P(c_i | z)$ directly obtained from the external knowledge base; $w(z) = \frac{\text{sim}(q, f(z))}{\sum_{z' \in Z_i} \text{sim}(q, f(z'))}$ is the normalized weight assigned based on retrieval similarity, where $z' \in Z_i$ iterates over all knowledge items in $Z_i$ to normalize the similarity scores (see detail in Appendix~\ref{appendix2}).

Notably, the value of $\lambda$ is dynamically determined by aggregating the retrieval weights $w(z)$, increasing linearly with their sum to emphasize external knowledge when retrievals are highly relevant. In practice, the value of $\lambda$ and the sum of $w(z)$ are considered equivalent by us. This design ensures that when external knowledge is highly relevant, the model relies more on external information; when external knowledge has lower relevance, it primarily relies on the internal Generator's predictions. 
The Fusionizer module employs a weighted distribution fusion strategy, leveraging Faiss to accelerate similarity computations, achieving an optimal balance between prediction accuracy and real-time performance. Faiss's efficient indexing structure ensures low-latency weighted fusion, achieving an optimal balance between computational efficiency and guessing precision.

\subsection{The way to generate passwords }
\label{subsec:password_generation}

The password generation process of KAPG proceeds iteratively, integrating the Generator, Retriever, and Fusionizer to construct password guesses step by step (see specific example instructions provided in Appendix~\ref{example generate}): 
1. \textbf{Initialization}: The process starts with a fixed prefix (e.g., \( c_1, c_2, c_3, c_4 = \texttt{\#\#\#\#} \)) to mark the beginning.
2. \textbf{Knowledge retrieval}: Given the current prefix \( c_{i-4}, \ldots, c_{i-1} \), the Retriever fetches relevant items \( Z_i \) from an external knowledge base via L2 similarity.
3. \textbf{Internal prediction}: The Generator computes the next-character distribution based on the prefix (i.e., \(P_{\text{int}}(c_i \mid c_{i-4}, \ldots, c_{i-1})\)).   
4.\textbf{Fusion}: The Retriever aggregates external distributions (i.e., \(P_{\text{ext}}(c_i \mid Z_i)\). The Fusionizer combines the Generator’s
and Retriever’s distributions by \(P(c_i \mid c_{i-4}, \ldots, c_{i-1}, Z_i) = (1 - \lambda) P_{\text{int}} + \lambda P_{\text{ext}}\).
5. \textbf{Sampling and update}: A character is sampled from the fused distribution and appended to the prefix (forming \( c_{i-4}, \ldots, c_{i-1} \)). The prefix length is capped at 4 characters. Excessively long prefixes may cause the Retriever to return overly specific knowledge items, reducing the Fusionizer module's ability to generate varied password patterns. 
6. \textbf{Termination}: Steps repeat until a valid password is generated, meeting a minimum length (e.g., 5 characters) or reaching the termination symbol (e.g., ``\textbackslash n'').

Notably, we provide a theoretical analysis of the validity of KAPG in Appendix~\ref{appendix3}, covering its information capacity, generalization error, knowledge integration advantages, and the rationality of using simple generators.

\section{Evaluation}
\label{5}
Here, we evaluate the guessing ability of KAPG\footnote{To ensure transparency and reproducibility, we have released the full code of KAPG along with the external knowledge base on anonymous.4open.science. The repository is available at: https://anonymous.4open.science/r/RAGGuess-CFE3} in multiple guessing scenarios consisting of several datasets, compared to state-of-the-art models. Additionally, we analyse the overlap rate of cracked passwords and model efficiency of KAPG against other models.
\subsection{Setup}
\subsubsection{Experimental scenarios}
To evaluate the guessing capabilities of KAPG and baseline models, we designed two experimental scenarios (intra-site and cross-site guessing) based on twelve breached password datasets. The intra-site guessing scenario (e.g., Linkedin $\rightarrow$ Linkedin) simulates a realistic situation where a single website experiences multiple password leaks~\cite{wang2023password}, making it an effective testbed for assessing a model’s ability to learn and replicate consistent password patterns within the same site. Specifically, in this scenario, both the training and test sets were drawn from the same breached dataset. To examine the models’ performance under various data scales, we trained them using training sets of two different sizes (i.e., 1.8 and 0.18 million entries, denoted as 1.8M and 0.18M). In contrast, the cross-site guessing scenario (e.g., Dodonew $\rightarrow$ Rockyou) reflects a more practical attack context, where attackers lack access to previously leaked passwords from the target site. To emulate this, we utilized datasets from different sources for the training (1.8M and 0.18M) and test sets, enabling us to evaluate the models’ generalization capabilities. In both cases, we set the size of the test set to 200,000 entries. Notably, following the recommendation in~\cite{wang2023password}, we did not remove duplicate password pairs appearing in the test sets.

\subsubsection{Model configuration}
We compared KAPG against state-of-the-art password guessing models, including PCFG, OMEN, RFGuess, FLA, and CKL\_PCFG. The configurations for each model are detailed below.

\textbf{PCFG}~\cite{weir2009password}. We employed the latest PCFG model (version 4.6~\cite{pcfg}), which incorporates an expanded set of password substring categories, such as keyboard patterns, website-specific patterns, and email patterns. Its parameter settings adhered to the recommendations provided in the open-source code. 
\begin{figure*}[htbp]
        \centering
	\begin{subfigure}{0.24\linewidth}
		\centering
		\includegraphics[width=1\linewidth]{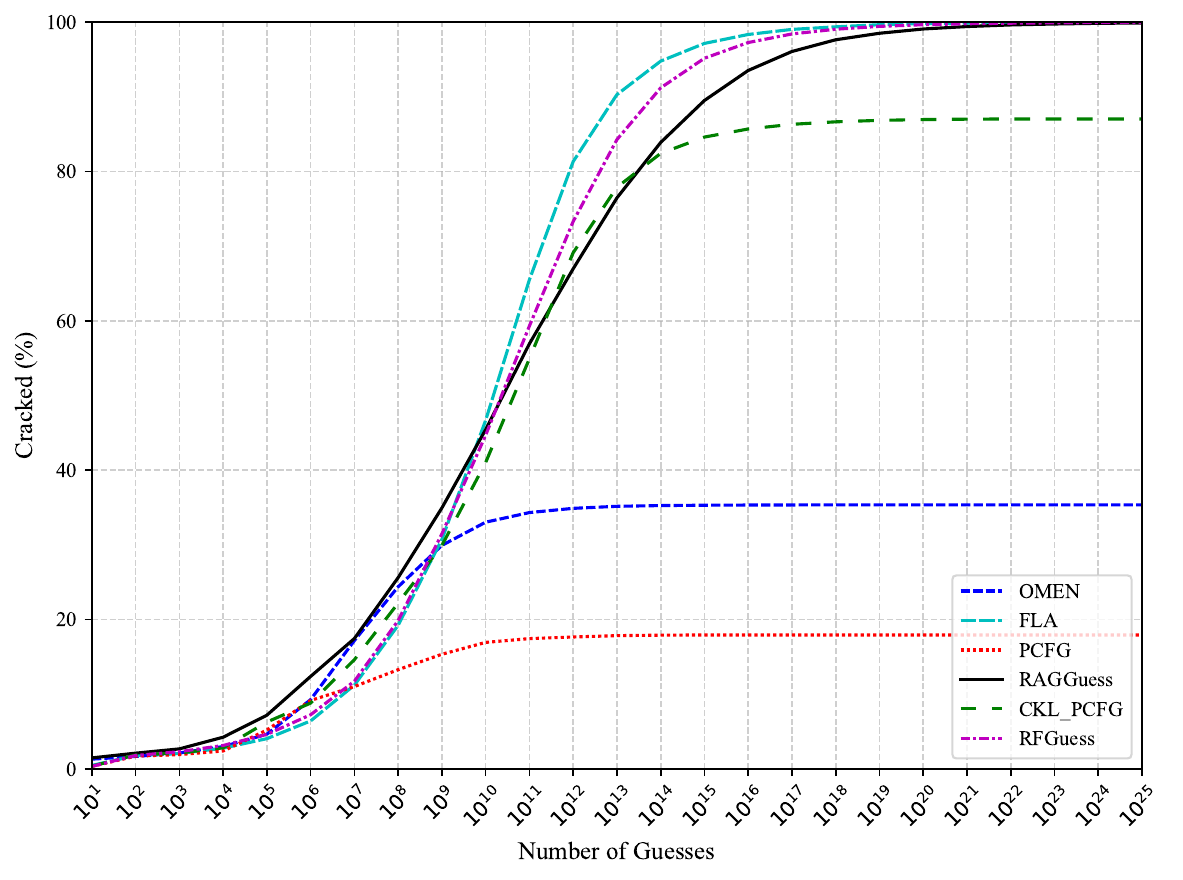}
		\caption{\textmd{0.18M CSDN $\rightarrow$ Linkedin}}
		\label{CSDN_length}
	\end{subfigure}
        \centering
	\begin{subfigure}{0.24\linewidth}
		\centering
		\includegraphics[width=1\linewidth]{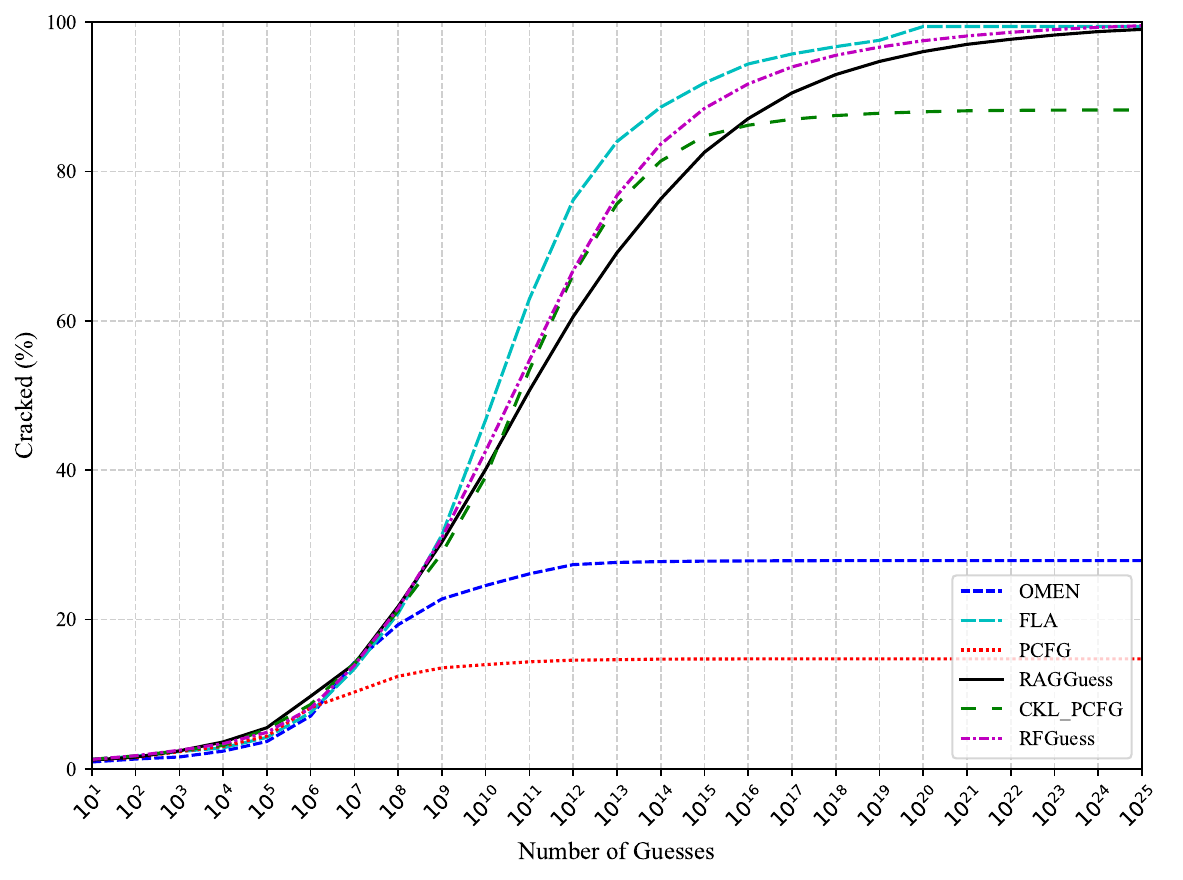}
		\caption{\textmd{0.18M Dodonew $\rightarrow$ Rockyou}}
		\label{CSDN_length}
	\end{subfigure}
        \centering
	\begin{subfigure}{0.24\linewidth}
		\centering
		\includegraphics[width=1\linewidth]{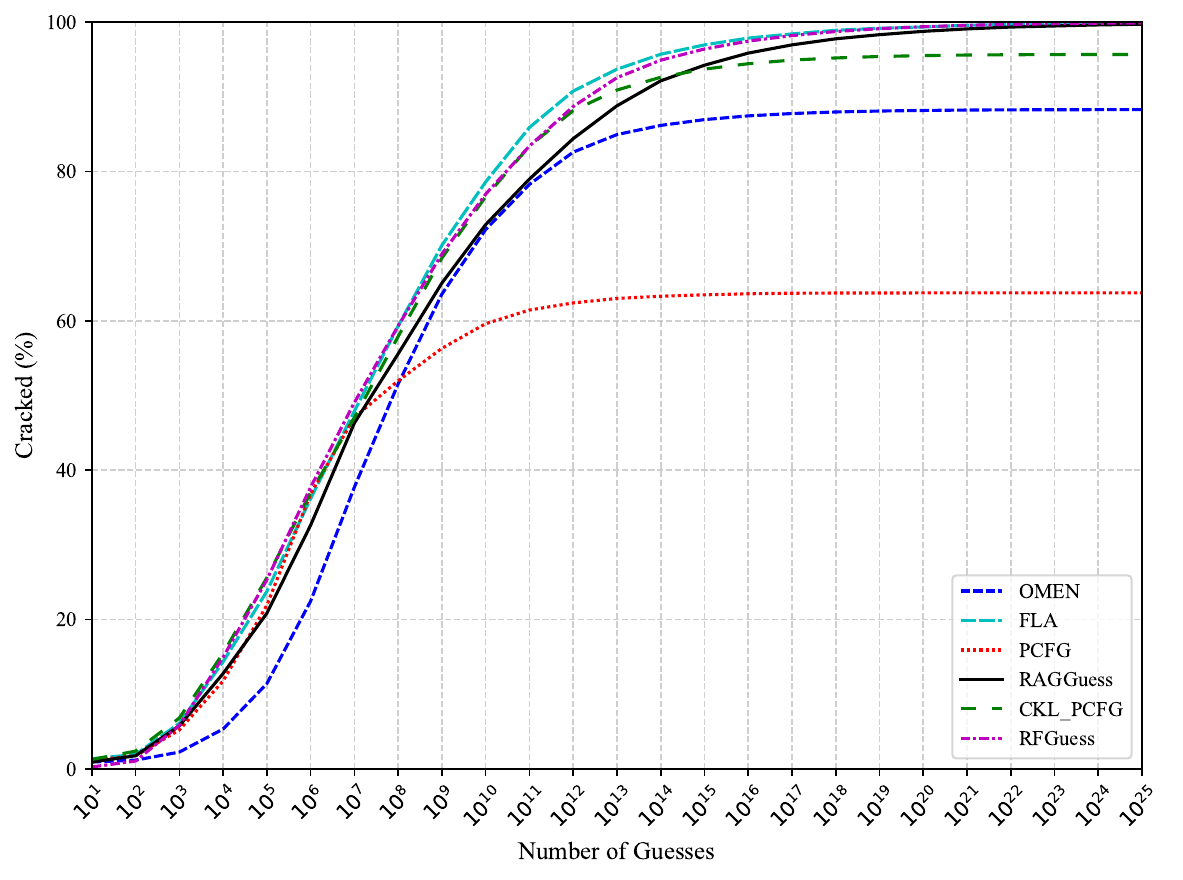}
		\caption{\textmd{1.8M Neopets $\rightarrow$ Rockyou}}
		\label{CSDN_length}
	\end{subfigure}
        \centering
	\begin{subfigure}{0.24\linewidth}
		\centering
		\includegraphics[width=1\linewidth]{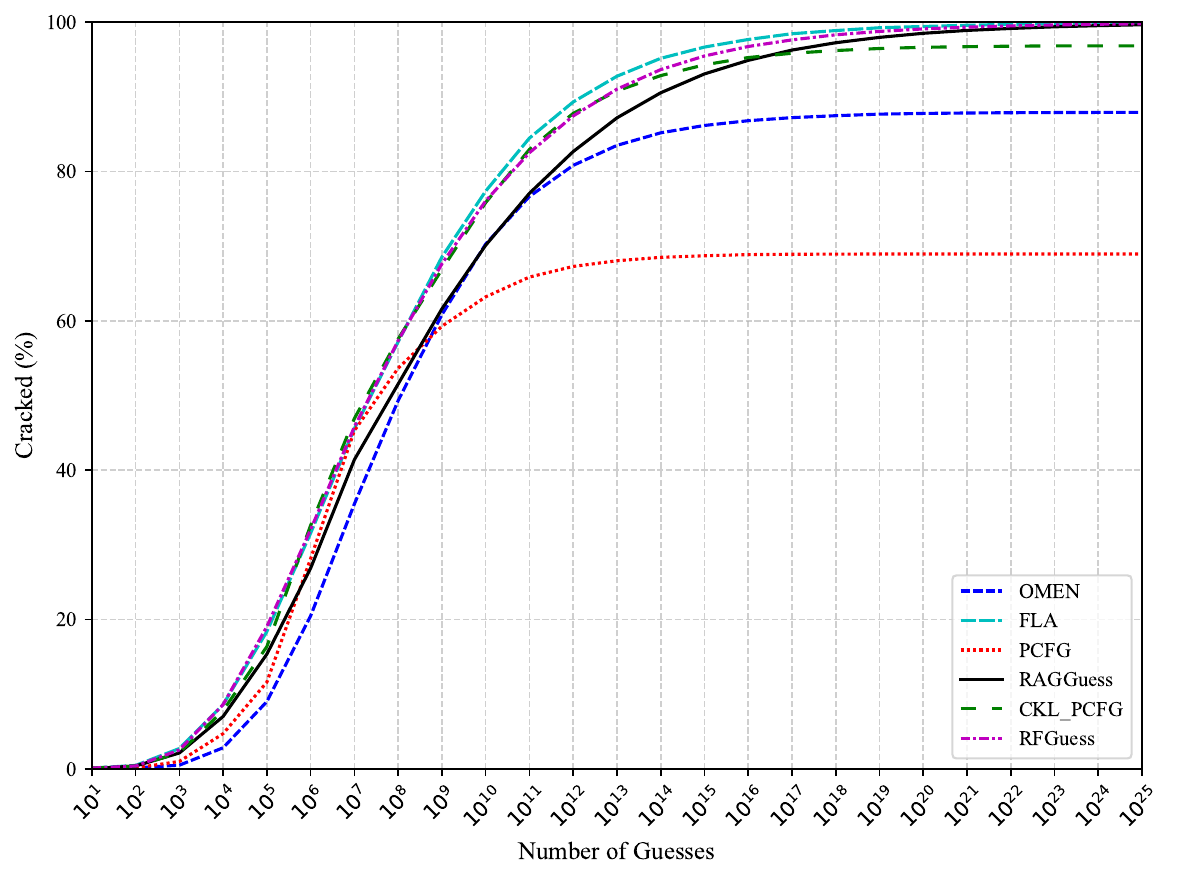}
		\caption{\textmd{1.8M Mathway $\rightarrow$ Neopets}}
		\label{CSDN_length}
	\end{subfigure}
        \centering
        \begin{subfigure}{0.24\linewidth}
            \centering
            \includegraphics[width=1\linewidth]{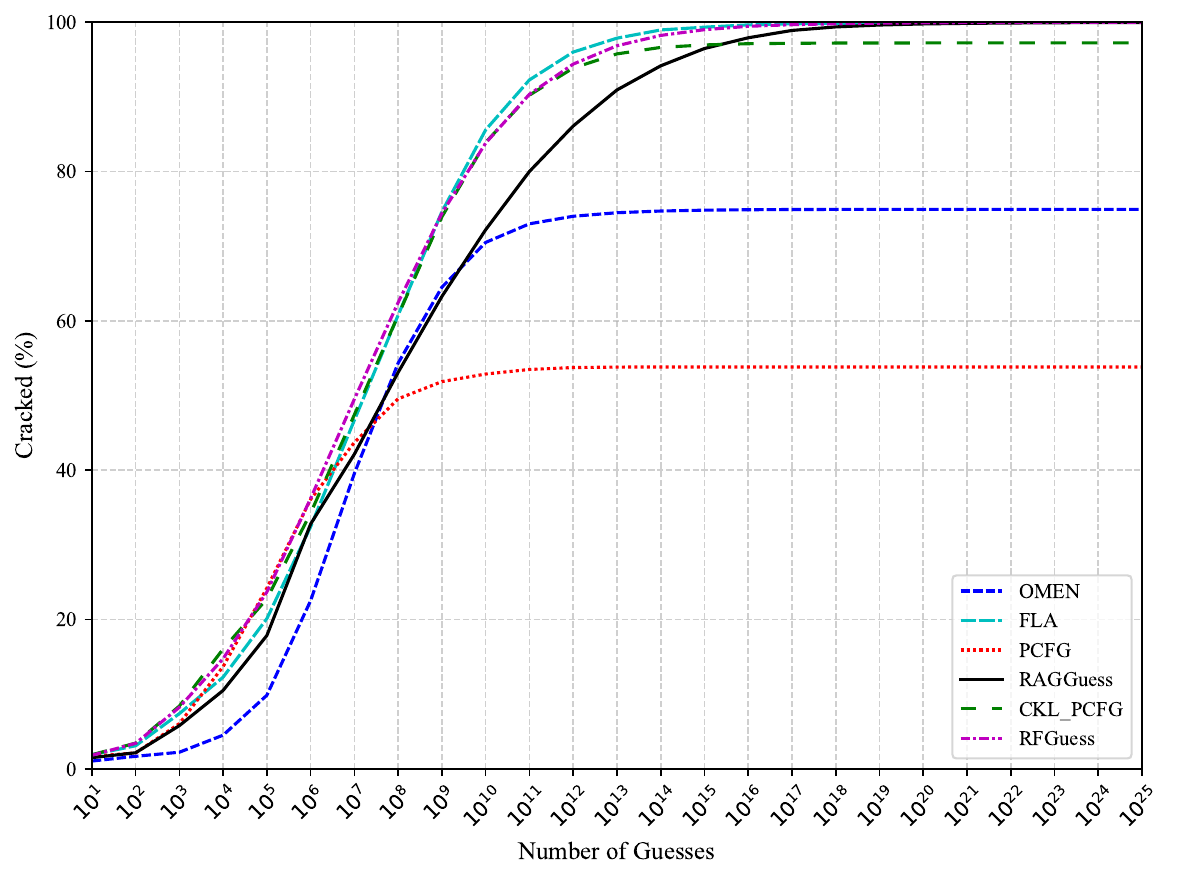}
            \caption{\textmd{0.18M Linkedin $\rightarrow$ Linkedin}}
            \label{CSDN_length}
        \end{subfigure}
        \centering
	\begin{subfigure}{0.24\linewidth}
		\centering
		\includegraphics[width=1\linewidth]{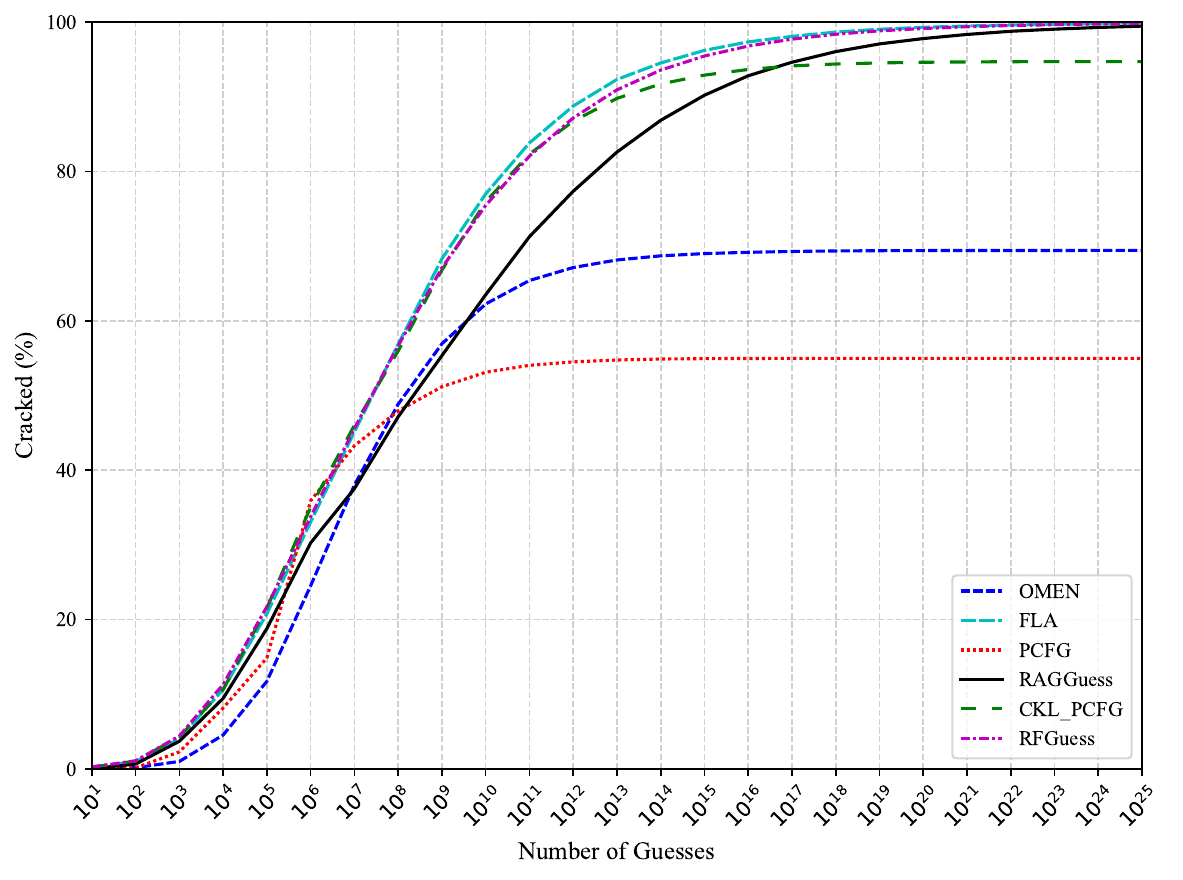}
		\caption{\textmd{0.18M Neopets $\rightarrow$ Neopets}}
		\label{CSDN_length}
	\end{subfigure}
        \centering
	\begin{subfigure}{0.24\linewidth}
		\centering
		\includegraphics[width=1\linewidth]{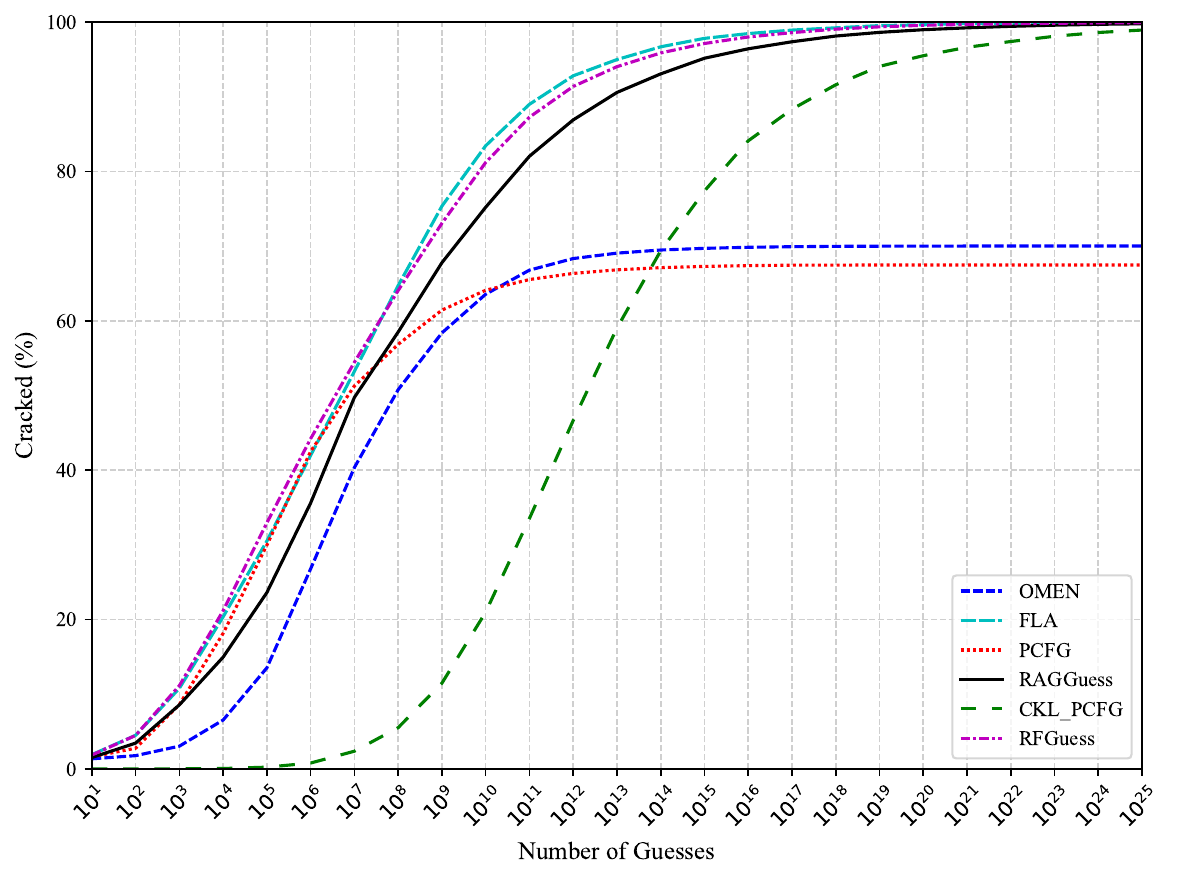}
		\caption{\textmd{1.8M Rockyou $\rightarrow$ Rockyou}}
		\label{CSDN_length}
	\end{subfigure}
        \centering
	\begin{subfigure}{0.24\linewidth}
		\centering
		\includegraphics[width=1\linewidth]{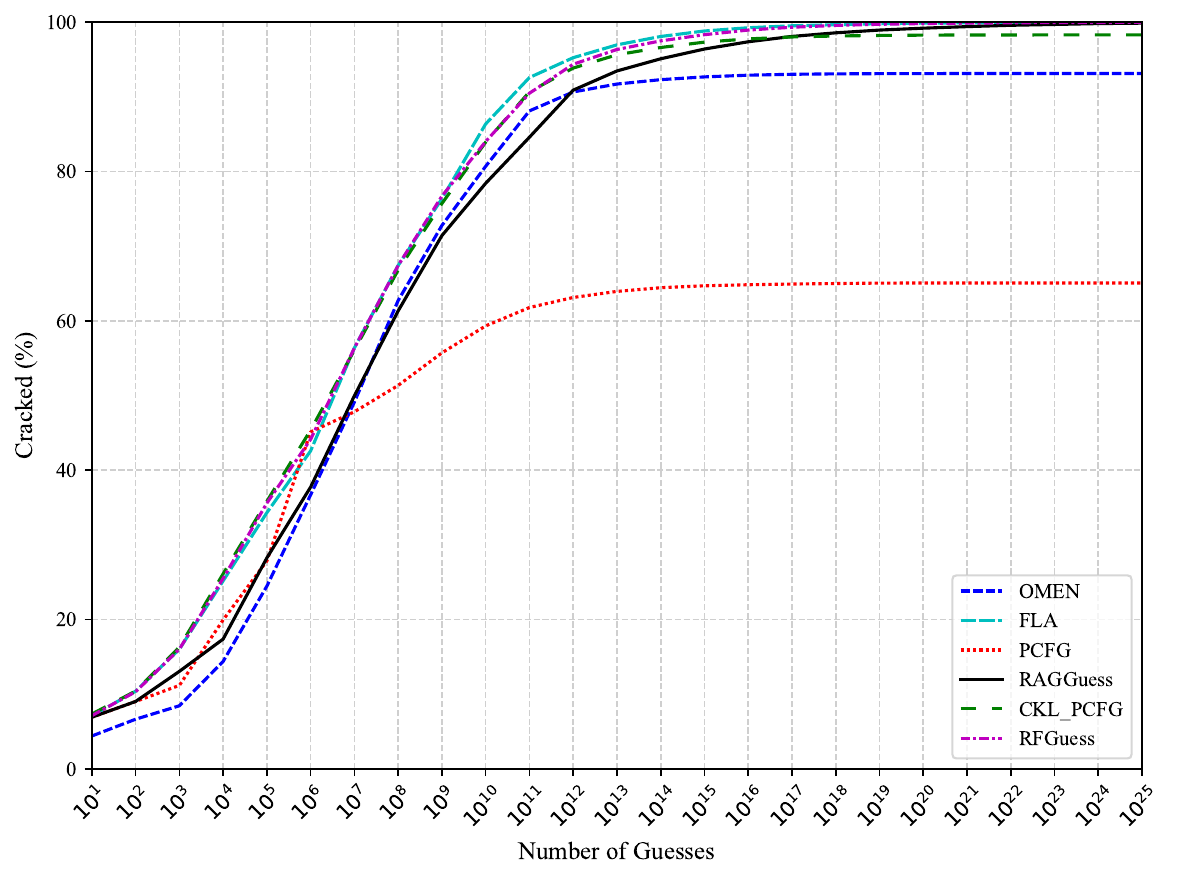}
		\caption{\textmd{1.8M 7K7K $\rightarrow$ 7K7K}}
		\label{CSDN_length}
	\end{subfigure}
	\caption{Guessing performance: PCFG, OMEN, FLA, RFGuess, CKL\_PCFG and KAPG in intra-site and cross-site password guess.}
	\label{fig:guessing result}
\end{figure*}
\textbf{OMEN}~\cite{durmuth2015omen}. The Ordered Markov Enumerator, an enhanced Markov model, generates password guesses in descending order of probability. We adopted a 4-order OMEN model, with other parameters configured according to the open-source code’s recommendations~\cite{omen}. 
\textbf{RFGuess}~\cite{wang2023password}. This random forest-based password guessing model was configured with 30 decision trees, each having a minimum of 10 leaf nodes and a maximum feature ratio of 80$\%$, as recommended by~\cite{wang2023password}. Remaining parameters followed the default settings of the scikit-learn framework. 
\textbf{FLA}~\cite{melicher2016fast}. A password guessing model based on LSTM, FLA was configured with three LSTM layers (200 units each) and two fully connected layers, trained for 20 epochs, following the parameter settings outlined in~\cite{fla}. 
\textbf{CKL\_PCFG}~\cite{xu2021chunk}. This chunk-level PCFG model diverges from traditional PCFG by using the BPE algorithm~\cite{gage1994new} to segment passwords and dynamically defining substring categories based on the number of character types within each chunk. Its configuration adhered to the recommendations in~\cite{xu2021chunk}. 
\textbf{KAPG}. Our proposed knowledge-augmented password guessing model was configured with a prefix length limit of 4 characters. The retriever module utilized a Faiss-based vector store with a top-\( k \) retrieval setting of \( k = 10 \). Generator employs a Markov-based approach to model the probability distribution of password sequences. 


\subsection{Result}
Directly generating password guesses and calculating their match rate with the test set (i.e., the cracking rate) poses significant computational challenges, particularly when the number of guesses becomes exceedingly large. To address this issue and accurately evaluate the cracking performance of each model, we employed the Monte Carlo algorithm~\cite{dell2015monte} to approximate the number of guesses required by each model to crack passwords in the test set. This approach, widely validated and adopted in numerous password guessing studies~\cite{wang2023password,xu2021chunk,melicher2016fast}, effectively balances computational efficiency with result accuracy. We show some of the guessing results in Figure~\ref{fig:guessing result}, and additional results can be found in Figure~\ref{fig:supplementary guessing result} in the Appendix.

\subsubsection{Intra-site}
Figure~\ref{fig:guessing result} illustrates the guessing performance of various models in intra-site scenarios with training datasets of 0.18M and 1.8M passwords. KAPG consistently demonstrates superior cracking rates across all intra-site scenarios, outperforming other models. Specifically, under the 0.18M training data size, at a benchmark of $10^{25}$ guesses, KAPG achieves cracking rate improvements of 113.5$\%$ (from 81.6\% to 247.4\%), 38.9$\%$ (from 25.9\% to 57.5\%), and 4.1$\%$ (from 2.8\% to 6.3\%) over PCFG, OMEN, and CKL\_PCFG, respectively. Compared to the state-of-the-art models FLA and RFGuess, KAPG exhibits comparable cracking rates, indicating similar predictive capabilities. For the 1.8M training data size, at the same $10^{25}$ guess benchmark, KAPG improves cracking rates by 50.6$\%$ (from 47.9\% to 87.9\%), 22.4$\%$ (from 7.3\% to 42.5\%), and 1.3$\%$ (from 1.1\% to 3.9\%) over PCFG, OMEN, and CKL\_PCFG, respectively. Against FLA and RFGuess, KAPG maintains equivalent cracking rates, underscoring its robust performance even with larger training datasets.

\subsubsection{Cross-site}
Figure~\ref{fig:guessing result} presents the guessing performance of different models in cross-site scenarios. Overall, KAPG achieves higher cracking rates across all cross-site scenarios, demonstrating strong generalization capabilities. 
Under the 0.18M training data size, at a benchmark of $10^{25}$ guesses, KAPG outperforms PCFG, OMEN, and CKL\_PCFG with cracking rate improvements of 283.8$\%$ (from 166.1\% to 457.0\%), 92.2$\%$ (from 56.0\% to 255.1\%), and 7.4$\%$ (from 2.4\% to 14.8\%), respectively. For the 1.8M training data size, at the same $10^{25}$ guess benchmark, KAPG surpasses PCFG, OMEN, and CKL\_PCFG by 100.7$\%$ (from 57.6\% to 187.2\%), 29.6$\%$ (from 13.3\% to 55.4\%), and 4.1$\%$ (from 2.9\% to 4.3\%), respectively. When compared to the leading models FLA and RFGuess, KAPG achieves equivalent cracking rates across both training data sizes, highlighting its competitive performance in cross-site scenarios.

Notably, as the training data size increases from 0.18M to 1.8M, most models exhibit significant improvements in cracking rates, whereas KAPG shows a relatively modest increase, indicating a lower dependency on training data scale. Even with a smaller training dataset (0.18M), KAPG demonstrates exceptional cracking performance. This advantage stems from KAPG’s design strengths: By integrating external knowledge retrieval with dynamic generation capabilities, KAPG efficiently captures password patterns, ensuring robust performance even in data-constrained scenarios.

\subsection{Analysis}
The multiple guessing scenarios designed above validate the excellent cracking performance of KAPG. However, focusing only on the cracking rate cannot fully reveal the characteristics of the model and its potential in practical applications. Therefore, we further analyse the performance of KAPG with comparative models in terms of cracking password overlap rate and model efficiency (including model size, training time and generation speed).

\subsubsection{Overlap}

\begin{figure*}[htbp]
	\centering
	\begin{subfigure}{0.49\linewidth}
		\centering
		\includegraphics[width=1\linewidth]{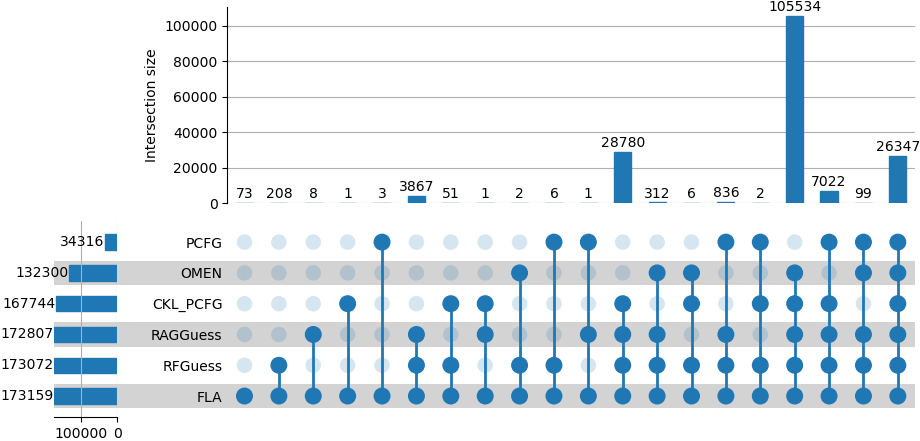}
		\caption{\textmd{Netease}}
		\label{CSDN_length}
	\end{subfigure}
	\centering
	\begin{subfigure}{0.49\linewidth}
		\centering
		\includegraphics[width=1\linewidth]{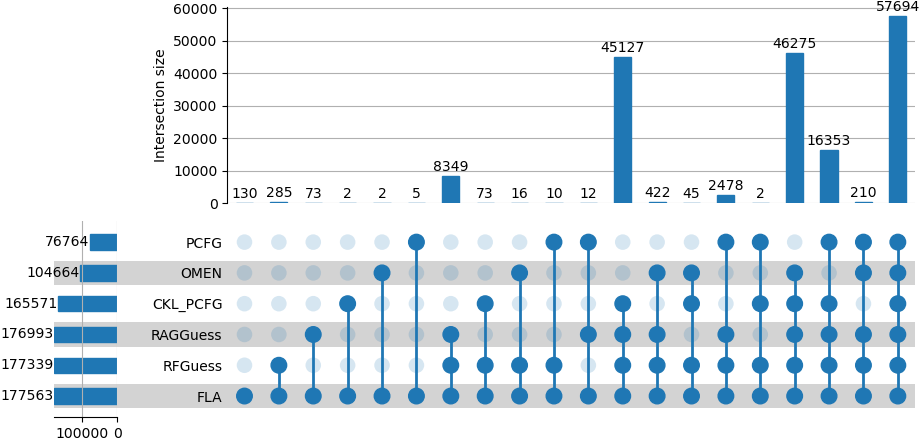}
		\caption{\textmd{Mathway}}
		\label{Rockyou_length}
	\end{subfigure}
	\caption{Overlap analysis of password cracking.}
	\label{fig:overlap}
\end{figure*}

We calculated the overlap rates of guessed passwords generated by various models across different datasets (see Figure~\ref{fig:overlap} exemplified by Netease and Rockyou) to assess the uniqueness and redundancy in their password guessing processes. For the Netease dataset, PCFG, OMEN, CKL\_PCFG, RFGuess, FLA, and KAPG collectively cracked 173,159 passwords, of which 26,347 were successfully identified by all models. Notably, FLA uniquely cracked 73 passwords that other models failed to identify, while FLA and KAPG together exclusively cracked an additional 208 passwords, highlighting KAPG’s distinct cracking capability. In the Rockyou dataset, the six models collectively cracked 177,563 passwords, with 57,694 being commonly identified across all. Similarly, FLA uniquely cracked 130 passwords, and in conjunction with KAPG, they exclusively cracked 285 additional passwords, further underscoring KAPG’s unique cracking prowess across datasets.

Despite KAPG’s demonstrated uniqueness, there remains room for improvement compared to other models. Specifically, on the Netease and Rockyou datasets, KAPG failed to crack 99 and 210 passwords, respectively, that were successfully identified by the other models. This discrepancy suggests that, while KAPG excels in exclusively cracking passwords, its coverage could be enhanced through design optimization to reduce blind spots in overlap with existing models.

\subsubsection{Model efficiency}
We quantitatively evaluated the efficiency of each model across three dimensions—model size, training time, and generation speed—with the results presented in Table~\ref{tab:model_efficiency}. KAPG has a model size of 1,091.2 MB, which corresponds to approximately 78.8$\%$ of RFGuess (1,384.3 MB) and 10.8$\%$ of OMEN (10,035.2 MB), indicating a relatively modest storage footprint compared to these models. However, KAPG’s model size remains substantially larger than that of PCFG (37.8 MB), FLA (5.5 MB), and CKL\_PCFG (59.6 MB). This increase in size is primarily attributed to the integration of an external knowledge base, which enhances its password guessing capabilities. While this design introduces a trade-off in storage requirements, it significantly boosts KAPG’s performance, making it viable even in resource-constrained environments.

Regarding training time, KAPG achieves an average of 64.3 seconds, outperforming all baseline models except OMEN (5.3 seconds), including PCFG (408.3 seconds), FLA (103,010.1 seconds), RFGuess (624.3 seconds), and CKL\_PCFG (19,291.1 seconds). Despite its relatively large model size, KAPG achieves a short training duration, making it highly efficient to train. This efficiency stems from the fact that storage capacity, which supports the large model size, is generally easier and less costly to scale than computational power, which is critical for training speed. Consequently, KAPG is particularly well-suited for deployment on devices with limited computational resources, such as edge devices or embedded systems. Nevertheless, compared to OMEN (c-coded), there remains room for improvement, and future optimizations could focus on streamlining the external knowledge retrieval module to further reduce training time.

In terms of generation speed, KAPG produces approximately 588,144.2 guesses per second, significantly outperforming PCFG (207,921.1 guesses/second), FLA (3,625.2 guesses/second), RFGuess (93.6 guesses/second), and CKL\_PCFG (302,793.1 guesses/second), demonstrating strong generation efficiency. However, due to the need for external information retrieval during guess generation, KAPG’s speed lags behind OMEN, the current leader at 1,679,402.1 guesses per second. To address this bottleneck, future improvements could focus on optimizing the retrieval algorithm or implementing caching mechanisms to minimize latency in accessing external information. 
Overall, KAPG achieves a commendable balance between model efficiency and cracking performance.

\begin{table}[!htbp]\centering
\setlength\tabcolsep{6pt}
  \caption{The model efficiency of PCFG, OMEN, FLA, RFGuess, CKL\_PCFG and KAPG.}
  \label{tab:model_efficiency}
  \resizebox{\columnwidth}{!}{
  \begin{tabular}{cccc}
  \toprule
  Models& Model size & Training time & Generated pwd/s \\
  \midrule
  \texttt \textmd{PCFG~\cite{pcfg}}&37.8 MB&408.3 s&207,912.1\\
  
  \texttt \textmd{OMEN~\cite{durmuth2015omen}}&10,035.2 MB&5.3 s&1,679,402.1\\
  
  \texttt \textmd{FLA~\cite{melicher2016fast}}&5.5 MB&103,010.1 s&3,625.2\\

  \texttt \textmd{RFGuess~\cite{wang2023password}}&1,384.3 MB&624.3 s&93.6\\

  \texttt \textmd{CKL\_PCFG~\cite{wang2023password}}&59.6 MB&19,291.1 s&302,793.1\\

  \texttt \textmd{KAPG}&1,091.2 MB&64.3 s&588,144.2\\

  \bottomrule
  \end{tabular}}
\end{table}

\section{A security application: password strength meter}
Password strength meters (PSMs)~\cite{pasquini2020interpretable, xu2021chunk,dlowe2016zxcvbn, guo2018lpse} are tools integrated into websites or applications to provide real-time feedback on the guessability of a user's password. The primary objective of PSMs is to quantify the difficulty of guessing a password, thereby guiding users toward more secure choices. Existing PSMs can be broadly categorized into two types: rule-based and data-driven. Rule-based PSMs evaluate password strength using predefined rule, such as password length, character types (uppercase and lowercase letters, digits, special characters), common patterns (e.g., keyboard sequences or repeated characters), and weak password blacklists. Each rule is assigned a fixed score, and the cumulative score is mapped to a strength level. Representative works include NIST-PSM~\cite{NIST1} and Zxcvbn~\cite{dlowe2016zxcvbn}. Data-driven PSMs, in contrast, leverage models trained on real-world leaked password datasets to estimate the likelihood of a password being generated, which is then used to infer guesses required to crack it. A common approach involves Monte Carlo methods to establish the relationship between password probability and guessing attempts. Representative works include Markov-~\cite{castelluccia2012adaptive}, PCFG-~\cite{houshmand2012building}, FLA-~\cite{melicher2016fast}, and CKL\_PCFG-based PSMs~\cite{xu2021chunk}. However, user's password preferences are influenced by dynamic factors, such as popular culture (e.g., movies, social media trends), seasonal events (e.g., holidays, trending topics), and community-specific conventions (e.g., naming patterns among gamers or developers). These factors cause password patterns to evolve significantly over time and across contexts. Both rule-based and data-driven PSMs typically rely on static rules or models constructed offline during a cold start. Without periodic large-scale retraining, these systems struggle to adapt to emerging keywords or community-specific patterns. Furthermore, user populations across different websites or industries exhibit distinct linguistic habits and topical preferences, rendering a one-size-fits-all PSM suboptimal and often misaligned with specific contexts. These limitations lead to frequent misestimations of password strength in real-world deployments, potentially overestimating the security of weak passwords or underestimating the robustness of strong ones. More critically, such misjudgments may inadvertently encourage users to adopt password patterns that superficially meet PSM criteria but are, in practice, insecure.

\begin{figure}[!htbp]
    \centering
    \includegraphics[scale=0.3]{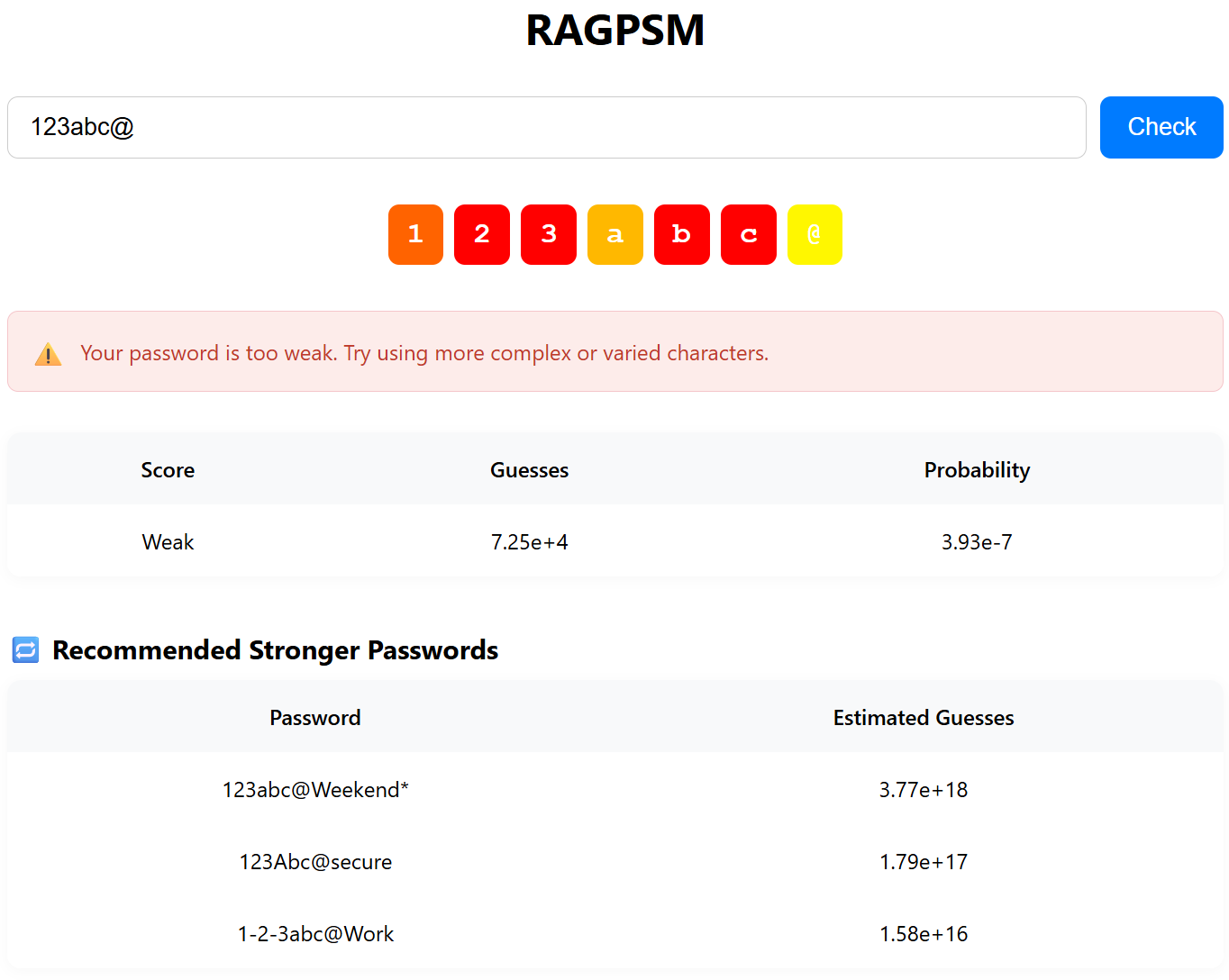}
    \caption{Interface of KAPSM.}
    \label{fig:ragpsm interface}
\end{figure}

A strong PSM should exhibit two critical capabilities: \textbf{Context adaptation}. The capacity to provide tailored evaluations based on linguistic backgrounds, industry characteristics, or community-specific cultures. For example, the password \texttt{woaini520} should be identified as easily guessable in Chinese-language contexts. \textbf{Extensible updating}. The ability to continuously incorporate new knowledge and adapt to evolving password distributions at a given site. Against this backdrop, KAPG inherently possesses a retrieval-enhanced structure that can integrate external semantic information and perceive trends in user behaviour, thereby providing a model foundation for constructing a context-aware, dynamically adaptable PSM. Based on this, we further constructed a KAPG-based password strength meter KAPSM. The Algorithm~\ref{alg:ragpsm} in the Appendix provides the working principle of KAPSM. 
Furthermore, we provide the implementation details of KAPG. 
We deployed KAPSM on a remote server and developed a corresponding web front-end that allows users to interactively access it through a browser (Figure~\ref{fig:ragpsm interface} shows the KAPSM interface, which can be accessed at https://4a5d38ac04cf.ngrok-free.app). This web front-end primarily includes the following functional modules. 
Character-level strength evaluation: The system colors each character of the password entered by the user to reflect its local security. We assign red to the highest character probability and yellow to the lowest, with the color changing linearly based on the character probability. 
Overall password strength evaluation: The system provides the predicted probability of the password and the estimated number of guesses required to crack it, and outputs an overall password strength rating (e.g., weak, medium, or strong). Specifically, if the number of guesses is less than $10^7$, the password is considered weak; if it is between $10^7$ and $10^{14}$, the password is considered medium; and if it is greater than $10^{14}$, the password is considered strong. 
Recommended password generation: After the user enters a password, the system uses a Large Language Model (such as DeepSeek) to generate multiple stronger candidate passwords. These recommendations enhance security while preserving the original password's memorability as much as possible. 
The backend system runs the complete KAPSM inference process. Its core components include prefix encoding, character-level prediction generation, and approximate vector retrieval combined with FAISS for external knowledge fusion. 
To enable the model to continuously adapt to user password evolution after deployment, KAPSM borrows the dynamic password guessing mechanism. Specifically, after being deployed to an actual website, KAPSM automatically receives new passwords submitted by users during registration or modification and embeds them into the external knowledge base for updating. With the continuous accumulation of real password samples, KAPSM can gradually adjust its prediction distribution for the next character, thereby adapt to the password construction habits of current website users. This mechanism enables KAPSM to achieve customised evaluation capabilities for different websites, reflecting the security strength of each password in a specific usage environment more realistically.

\begin{table}[!htbp]\centering
\setlength\tabcolsep{6pt}
  \caption{Performance comparison of PSMs.}
  \label{tab:rank correlation}
  \resizebox{.45\textwidth}{!}{
  \fontsize{20pt}{23pt}\selectfont 
  \begin{tabular}{ccccccc}
  \toprule
  Models & Accuracy & Response time & Required disk space\\
  \midrule
  FLA~\cite{melicher2016fast} & 0.39 & 104.5 ms & 5.5 MB\\
  PCFG~\cite{pcfg} & 0.35 & 0.2 ms & 37.8 MB\\
  CKL\_PCFG~\cite{xu2021chunk} & 0.47 & 23.0 ms & 59.6 MB\\
  RFGuess~\cite{wang2023password} & 0.45 & 19.1 ms & 1,331.2 MB\\
  OMEN~\cite{durmuth2015omen} & 0.41 & 0.03 ms & 10,035.2 MB\\
  KAPSM & 0.53 & 0.7 ms & 1,091.2 MB\\
  \bottomrule
  \end{tabular}}
\end{table}

We tested KAPSM's ability to evaluate password strength using accuracy (using 12 datasets). Following the guidelines defined by Golla and D{\"u}rmuth et al.~\cite{golla2018accuracy}, we used the weighted Spearman correlation coefficient to calculate the correlation between KAPSM's password strength assessment ranking and the password frequency ranking (values range from -1 to 1, with 1 representing complete correlation and -1 representing complete independence). As shown in Table~\ref{tab:rank correlation}, KAPSM had the highest password strength assessment accuracy (FLA: 0.39, PCFG: 0.35, CKL\_PCFG: 0.47, RFGuess: 0.45, OMEN: 0.41, KAPSM: 0.53), and its feedback speed for password evaluation results is also almost the fastest (FLA: 104.5 ms, PCFG: 0.2 ms, CKL\_PCFG: 23.0 ms, RFGuess: 19.1 ms, OMEN: 0.03 ms, KAPSM: 0.7 ms). Evidence shows that a response time of less than 100 ms is acceptable to humans~\cite{nielsen1994usability}. It should be acknowledged that KAPSM requires slightly more disk space (1,091.2 MB) compared to traditional models. However, considering the significant improvement in evaluation capabilities brought by its fusion generation and retrieval enhancement mechanisms, this overhead is fully acceptable in modern computing environments.


Despite verifying the performance of KAPSM on multiple password datasets, this is insufficient to comprehensively reflect KAPSM's customisable adaptability to specific website user behaviour in real-world deployment. To this end, we designed a simulation experiments to construct a more realistic deployment scenario. Specifically, we simulated KAPSM deployed on the English-language website Rockyou, but trained KAPSM on the Chinese password dataset Dodonew. At the beginning of the experiment, KAPSM's external knowledge base did not contain any password information from Rockyou users. This provided a clear baseline for testing the model's gradual adaptation to website characteristics from the training environment. During the experiment, we gradually injected real user passwords selected from the Rockyou dataset into KAPSM's external knowledge base to simulate data updates brought about by user registration. We evaluated the difference in model performance in the following two scenarios. No external adaptation (KAPSM): The external knowledge base does not contain password data from website (i.e., Rockyou) users; gradual adaptation (KAPSM-0.01, KAPSM-0.1, and KAPSM-1): 0.01, 0.1, and 1 million website user passwords were added to the external knowledge base, respectively. We measured the model's accuracy in assessing the strength of passwords in the Rockyou testset.

The experimental results show that as the number of deployed website password samples included in the external knowledge base continues to increase, the evaluation accuracy of KAPSM steadily improves, verifying its excellent website adaptation and dynamic optimisation capabilities. Specifically, in the initial state without any Rockyou data, KAPSM's accuracy rate has reached 0.42, significantly outperforming existing mainstream models (FLA at 0.31, PCFG at 0.36, OMEN at 0.34, RFGuess at 0.36, CKL\_PCFG at 0.32). These results fully demonstrate that even in the initial stage when there is a lack of understanding of the deployment site, KAPSM shows strong generalisation ability and robustness. On this basis, we further examined the evolution trend of KAPSM after user data injection: After injecting 0.01 million Rockyou user passwords, the accuracy rate increased to 0.43; after injecting 0.1 million entries, the accuracy rate reached 0.44; after injecting 1 million entries, the accuracy rate further improved to 0.46. These results systematically verify that KAPSM can achieve adaptive evolution of websites in real deployment environments by continuously absorbing user password, thereby providing more accurate and robust strength assessments. This continuous learning and migration capability is a core advantage that traditional static password strength meters cannot provide.

\section{Discussion}
KAPG or KAPSM relies on an external knowledge base that stores aggregated n-gram sequences along with their corresponding next-character probability distributions, rather than raw passwords or any reversible representations. This design fundamentally reduces the risk of sensitive information leakage and complies with password security best practices recommended by standards such as NIST SP 800-63B~\cite{NIST}, which explicitly prohibit storing user passwords in plaintext or reversible hashed forms. However, prior research has shown that even under certain conditions, such aggregated statistical representations may still leak partial information about the underlying password distribution. Specifically, when the knowledge base is constructed from a relatively small dataset or when the attacker possesses auxiliary prior knowledge, n-gram frequency data can reveal limited structural cues regarding users’ password habits. Castelluccia et al. first raised this concern in their NDSS'12 work~\cite{castelluccia2012adaptive}, analyzing residual information leakage from probability-perturbed n-gram distributions. To mitigate such risks, differential privacy mechanisms have been proposed to formally restrict information exposure in statistical password modeling. While privacy-preserving design is essential, KAPSM also faces a more immediate and practical threat: Poisoning attacks targeting its update mechanism. In deployment, KAPSM dynamically adapts to user password behaviors by incorporating newly submitted passwords during registration or password changes. Although this continuous self-updating enhances the accuracy of site-specific password strength evaluation, it simultaneously introduces a potential attack. Without effective constraints, adversaries can inject large volumes of malicious passwords, causing KAPSM to misclassify weak passwords as strong ones, thereby undermining its core security functionality and turning its personalization advantage into bias drift vulnerabilities. This issue underscores the urgent need for further research into the privacy and security of PSM, as discussed in works related to membership inference attacks~\cite{xu2025account}. Accordingly, we have initiated efforts to develop a systematic poisoning attack detection and defense framework leveraging adversarial learning mechanism.

\section{Conclusion}
\label{7}
This paper reveals the significant impact of popular trends and cultural factors on users' password creation habits through an analysis of the evolution of password datasets. Based on these insights, we introduce KAPG, a pioneering knowledge-augmented guessing framework that combines external knowledge with generative capabilities. Experimental evaluations in both intra-site and cross-site scenarios demonstrate that KAPG exhibits exceptional guessing capabilities while maintaining competitive model efficiency and low password overlap redundancy. 
To mitigate guessing attacks, we designed and implemented KAPSM. Compared to mainstream PSMs, KAPSM achieves the highest strength evaluation accuracy. Furthermore, KAPSM demonstrated strong adaptability in simulated deployment scenarios.
\bibliographystyle{plain}
\bibliography{RAGGuess_reference}

\appendices
\section{Dynamic password guessing}
\label{Avalanche cracking}

The practical situation is that user-created passwords are often not isolated. They exhibit patterns of reuse across sites, incremental modifications (e.g., from \texttt{alice2021} to \texttt{alice2024}), and so on. Traditional guessing models, reliant on fixed training datasets, fail to capitalize on these dependencies, treating each guess as an independent event and thus missing opportunities to leverage successful cracks for subsequent predictions. To overcome this problem, we proposed a dynamic password guessing method based on KAPG (referred to as KAPG-DPG), which significantly improves the cracking ability by triggering the cascade effect of password guessing. Upon successfully cracking a password, KAPG-DPG identifies and integrates it into the external knowledge base, refining the character probability distribution of related prefixes. This process initiates a chain reaction: A single successful crack increases the likelihood of guessing related passwords in subsequent iterations.

\subsection{Method}

In implementing the update mechanism for dynamic password guessing, we first tackle a critical challenge: Preventing the model from overfitting to specific password patterns due to immediate updates to the external knowledge base following each successful guess. For instance, if a newly cracked password (e.g., \texttt{alice2021}) is instantly incorporated into the knowledge base, the model may quickly bias its character probability distribution toward this pattern’s features (e.g., a letter-digit combination). This bias could lead subsequent guesses to cluster around variants such as \texttt{alice2024} or \texttt{alice2023}, shrinking the exploration space and diminishing the model’s ability to target other prevalent patterns (e.g., \texttt{password123} or \texttt{abc123}). Such overfitting not only reduces cracking efficiency but also undermines the adaptability to diverse and evolving user behaviors.

To mitigate this risk and minimize the impact of noise from individual cracks, we devise a batch update strategy based on logarithmic intervals. Specifically, we partition the guessing process into intervals defined by powers of 10, updating the external knowledge base only at the end of each interval, i.e., after \( 10^k \) guesses, where \( k \) is a positive integer representing the interval tier (e.g., \( 10, 100, 1000 \)). The update frequency is thus governed by:
\(
t_u = \lfloor \log_{10} (n) \rfloor + 1, \quad n = 1, 2, \ldots,
\)
where \( n \) is the cumulative number of guesses, and \( t_u \) denotes the next update point. At each update, the knowledge base aggregates patterns from all successfully cracked passwords within the interval, adjusting the conditional probability distribution \( P(c | p) \) of character \( c \) given prefix \( p \) as:
\(
P(c | p) = \frac{\sum_{i \in Crack} \mathbb{I}(p, c \in pwd_i) + \alpha}{\sum_{i \in Crack} \mathbb{I}(p \in pwd_i) + \alpha |cs|},
\)
where \( Crack \) is the set of cracked passwords in the batch, \( \mathbb{I}(\cdot) \) is an indicator function, \( cs \) is the character set, and \( \alpha \) is a smoothing parameter (e.g., Laplace smoothing with \( \alpha = 1 \)).

This logarithmic batching approach prevents overfitting by delaying updates until sufficient data accumulates, ensuring that the model captures a broader representation of password patterns rather than overfitting to outliers. The logarithmic progression of intervals balances update frequency with exploration: early intervals (e.g., \( 10\) guesses) allow rapid adaptation to initial successes, while later, larger intervals (e.g., \( 1000\) guesses) stabilize the model against noise as the guess count grows.

Having determined the update timing, we further design a probability update rule to ensure that updates to the knowledge base accurately reflect authentic user behavior patterns. A key observation is that passwords cracked with fewer guesses are more likely to reflect common user habits, whereas those requiring more guesses may result from accidental matches and are less representative of typical behavior. For example, if the password \texttt{n\#mgka123} is cracked only after $10^{10}$ guesses, its pattern may not be representative, and its influence on subsequent guesses should be appropriately suppressed. To this end, we devise a probability increment rule based on the number of guesses: passwords cracked with fewer guesses exert a greater influence on the knowledge base. Formally, for a password cracked after guesses attempts, its probability increment is defined as $1/(guesses+1)$. This rule ensures that more representative password patterns have a larger impact on the knowledge base. 
When updating the knowledge base, if a newly cracked password already exists or partially exists in the knowledge base, historical data must be fused with the new data. To prevent abrupt fluctuations in the probability distribution, we adopt the Exponential Moving Average (EMA) method to update the probability distribution in the knowledge base. EMA balances the relative importance of historical probabilities and new inputs through a smoothing factor $\beta$, with the update formula expressed as: $P_{new}=\beta \times P_{old} + (1-\beta) \times P_{input}$. Here, $P_{new}$ is the updated probability, $P_{old}$ is the original probability in the knowledge base, $P_{input}$ is the probability increment from the new input, and $\beta \in [0,1]$ controls the smoothing degree (set to 0.8 in this experiment). The EMA method effectively mitigates sharp data fluctuations through smooth updates, ensuring that the probability distribution in the knowledge base stably reflects the evolution of user behavior patterns.

\subsection{Result}
We conducted experiments on two password datasets, CSDN and Rockyou, comparing the cracking performance of KAPG with and without the dynamic password guessing mechanism enabled (see Table~\ref{tab:avalanche_csdn}). The experiments measure the number of passwords cracked at various guessing intervals, defined by powers of 10. These intervals align with the batch update strategy described in the above, allowing us to assess the cumulative impact of KAPG-DPG on cracking efficiency as the number of guesses increases.

\begin{table*}[!htbp]\centering
\setlength\tabcolsep{6pt}
  \caption{Cracking performance of KAPG and KAPG-DPG on CSDN and Rockyou.}
  \label{tab:avalanche_csdn}
  \resizebox{1.8\columnwidth}{!}{
  \begin{tabular}{ccccc}
  \toprule
  & KAPG(train by CSDN) & KAPG-DPG (train by CSDN) &KAPG(train by Rockyou)& KAPG-DPG (train by Rockyou) \\
  \midrule

  \texttt \textmd{$10^{1}$} &14,855&14,863&3,031&3,708\\
  \texttt \textmd{$10^{2}$} & 22,849 &25,759 &6,899&7,341\\
  \texttt \textmd{$10^{3}$} & 25,816 &28,563 &17,136&17,634\\
  \texttt \textmd{$10^{4}$}  & 31,000 & 32,696&29,927&29,841\\
  \texttt \textmd{$10^{5}$}  &44,628  &46,076  &47,228&47,344\\
  \texttt \textmd{$10^{6}$}  & 54,947 &56,655 &71,217&73,552\\
  \texttt \textmd{$10^{7}$} &65,822  & 66,443 &99,494&100,423\\
  \texttt \textmd{$10^{8}$}  &77,148  & 77,226 &117,002&117,606\\
  \texttt \textmd{$10^{9}$}  & 98,998 &100,731 &135,579&136,260\\
  \texttt \textmd{$10^{10}$} &  117,868 & 118,763 &150,435&150,743\\

  \bottomrule
  \end{tabular}}
\end{table*}

On the CSDN, KAPG-DPG consistently outperforms the baseline KAPG across all guessing intervals. At $10^1$ guesses, KAPG-DPG cracks 14,863 passwords compared to 14,855 for the baseline, a modest improvement of 0.05$\%$. However, the advantage becomes more pronounced as the number of guesses increases. At $10^2$ guesses, KAPG-DPG cracks 25,759 passwords, an improvement of 12.7$\%$ over the baseline's 22,849. By $10^{10}$ guesses, KAPG-DPG cracks 118,763 passwords, surpassing the baseline's 117,868 by 0.8$\%$. This trend indicates that KAPG-DPG effectively leverages early successful cracks to trigger cascades of related passwords, with the most significant gains observed in the mid-range intervals, where the knowledge base updates have accumulated sufficient patterns to amplify cracking efficiency. 
The Rockyou exhibits a similar trend, with KAPG-DPG demonstrating notable improvements, particularly in the early to mid-range intervals. At 10 guesses, KAPG-DPG cracks 3,708 passwords, a substantial 22.3$\%$ improvement over the baseline's 3,031. This early advantage highlights KAPG-DPG's ability to quickly adapt to common patterns in the dataset. At 100 , the improvement is 6.4$\%$ (7,341 vs. 6,899), and at 1000 , it is 2.9$\%$ (17,634 vs. 17,136). Notably, the relative improvement diminishes in later intervals, suggesting that as the number of guesses increases, the knowledge base becomes saturated with patterns, reducing the marginal benefit of additional updates.

\section{Theoretical and empirical analysis of generator model selection}
\label{lstm_selection}
Through in-depth theoretical analysis and experimental validation, we ultimately chose a Markov chain-based statistical model rather than an LSTM neural network as our base generator. This seemingly ``conservative'' choice actually embodies profound technical considerations.

\textbf{Sparsity as an advantage.} Traditional Markov models, when encountering n-grams that do not appear in the training data, produce explicit ``unknown'' signals, triggering fallback mechanisms or returning Extremely low probabilities. This explicit expression of uncertainty has unique value in the KAPG framework:

\begin{align*}
&P_{Markov}(c_i | c_{i-n:i-1}) \\
&= \begin{cases} 
P(c_i | c_{i-n:i-1}) & c_{i-n:i-1} \in \mathcal{C}_{train} \\
\alpha \cdot P_{backoff}(c_i | c_{i-n+1:i-1}) & c_{i-n:i-1} \notin \mathcal{C}_{train}, \text{backoff} \\
\epsilon & \text{otherwise}
\end{cases}
\end{align*}

Where $\mathcal{C}_{train}$ represents the set of contexts appearing in the training data, $P_{backoff}$ represents the probability of backing off to a lower-order model, $\alpha$ is the smoothing parameter and $\epsilon$  represents an extremely low probability. When encountering unseen n-grams, the system triggers a backoff mechanism to an (n-1) - order model, or returns an extremely low but non-zero probability value through smoothing techniques, thereby providing a clear intervention opportunity for the retrieval augmentation mechanism.

\textbf{The over-generalization problem of LSTM neural networks.} In contrast, LSTMs generate seemingly reasonable probability distributions even for rare or unseen input sequences through continuous hidden state representations and nonlinear transformations:
$$P_{LSTM}(c_i | h_{i-1}) = \text{softmax}(W_o \cdot \tanh(W_h \cdot h_{i-1} + b))$$

Where $h_{i-1} \in \mathbb{R}^d$ is the hidden state vector from the previous time step, encoding the historical information of the sequence; $W_h \in \mathbb{R}^{m \times d}$ and $W_o \in \mathbb{R}^{|V| \times m}$ are the weight matrices for the hidden layer and output layer respectively ($|V|$ is the vocabulary size, $m$ is the hidden layer dimension, and $d$ is the hidden state vector dimension); $b \in \mathbb{R}^m$ is the bias term. This ``full coverage'' characteristic is an advantage when used alone, but becomes a disadvantage in the KAPG framework. LSTMs always output a complete probability distribution, and even when their internal representations exhibit high uncertainty in modeling the current context, this uncertainty is hidden within continuous probability values, making it difficult for the Fusionizer module to accurately identify and utilize. When Markov triggers backoff, its uncertainty directly translates into assigning greater weight to external knowledge. However, LSTM's continuous output makes it difficult to provide such clear signals.

\begin{figure}[!htbp]
    \centering
    \includegraphics[scale=0.3]{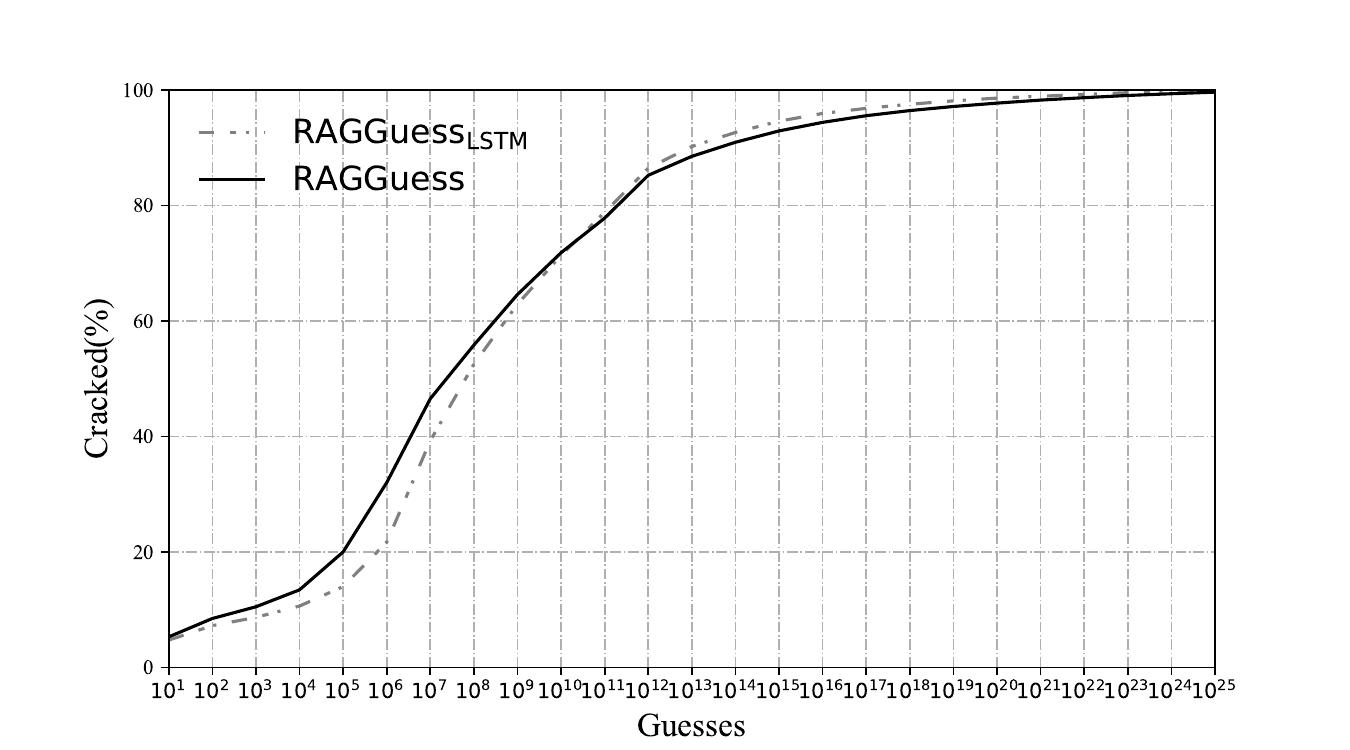}
    \caption{Password cracking performance comparison: KAPG vs. $\text{KAPG}_{LSTM}$.}
    \label{fig:lstm}
\end{figure}

Meanwhile, we compared the performance of both generators under the same framework. The experimental setup used a dataset of 1.8 million passwords from Dodonew for cross-site guessing on Tianya. The final experimental results (see Figure~\ref{fig:lstm}) show that KAPG combined with LSTM (called $\text{KAPG}_{LSTM}$) begins to exceed the cracking rate of Markov-based KAPG (called KAPG) starting from $10^{11}$ guesses. Under small guess numbers, KAPG actually demonstrates better guessing performance. KAPG training took only 64.3 seconds, while $\text{KAPG}_{LSTM}$ training took 3001.7 seconds. Additionally, KAPG generates 588,144.2 password guesses per second, while $\text{KAPG}_{LSTM}$ generates only 3.5 guesses per second.

\section{Correctness of the composite probability formula}
\label{appendix2}

In the KAPG framework, the Fusionizer module integrates the Generator’s internal predictions with the Retriever’s external knowledge to compute the probability of the next character \( c_i \) given the prefix \( c_{i-4}, \ldots, c_{i-1} \) and retrieved knowledge \( Z_i \). The composite probability is defined as: 
\(
P(c_i | c_{i-4}, \ldots, c_{i-1}, Z_i) = (1 - \lambda) P_{\text{int}}(c_i | c_{i-4}, \ldots, c_{i-1}) + \lambda P_{\text{ext}}(c_i | Z_i),
\)
where: \( P_{\text{int}}(c_i | c_{i-4}, \ldots, c_{i-1}) \) is the internal probability from the Generator’s 4th-order Markov model, based on static training data; \( P_{\text{ext}}(c_i | Z_i) = \sum_{z \in Z_i} w(z) P(c_i | z) \) is the external probability, with weights \( w(z) = \frac{\text{sim}(q, f(z))}{\sum_{z' \in Z_i} \text{sim}(q, f(z'))} \) based on similarity scores for the top-10 retrieved documents; \( \lambda \in [0, 1] \) is a dynamic weighting factor reflecting the relevance of external knowledge.

The sum over all possible \( c_i \) is: 
\(
\sum_{c_i} P(c_i | c_{i-4}, \ldots, c_{i-1}, Z_i) \\ =  \sum_{c_i} \left[ (1 - \lambda) P_{\text{int}}(c_i | c_{i-4}, \ldots, c_{i-1}) + \lambda P_{\text{ext}}(c_i | Z_i) \right].
\)
Since \( \sum_{c_i} P_{\text{int}}(c_i | c_{i-4}, \ldots, c_{i-1}) = 1 \) (Markov model) and:
\(
\sum_{c_i} P_{\text{ext}}(c_i | Z_i) = \sum_{c_i} \sum_{z \in Z_i} w(z) P(c_i | z) = \sum_{z \in Z_i} w(z) \sum_{c_i} P(c_i | z) = \sum_{z \in Z_i} w(z) \cdot 1 \approx 1,
\)
the composite probability sums to 1. 

We believe that similarity scores follow the Zipf distribution (as external knowledge and passwords created by humans satisfy the Zipf distribution), and these top 10 related results will account for a significant portion of the overall results.

\section{Example of password generated by KAPG}
\label{example generate}
To illustrate how KAPG operates, we present a concrete example of its iterative generation process:

\begin{enumerate}
    \item \textbf{Initialization}: Suppose the generation begins with a prefix (e.g., \texttt{love}).
    
    \item \textbf{Retrieval of external knowledge}: Given the current prefix, the Retriever identifies semantically aligned items from the external knowledge base, such as \texttt{lover}, \texttt{loved}, and \texttt{lovely}.
    
    \item \textbf{Internal prediction}: The Generator computes the \(P_{\text{int}}(c_i \mid c_{i-4}, \ldots, c_{i-1})\) to obtain a probability distribution\{ \texttt{y}: 0.02, \texttt{u}: 0.03, \texttt{1}: 0.01, \ldots \}.
    
    \item \textbf{Fusion with external knowledge}: The Retriever computes similarity scores for the retrieved items (e.g., 0.7 for \texttt{lover}, 0.8 for \texttt{loved}, 0.9 for \texttt{lovely}), yielding normalized weights \( w(z) \approx \{ 0.28, 0.32, 0.36 \} \). These contribute to the external distribution:
    \(
    P_{\text{ext}}(c_i \mid Z_i) = \{ \texttt{r}: 0.25, \texttt{d}: 0.30, \texttt{l}: 0.35, \ldots \}.
    \)
    The Fusionizer integrates the internal and external predictions as:
    \(
    P(c_i \mid c_{i-4}, \ldots, c_{i-1}, Z_i) = (1 - \lambda) P_{\text{int}}(c_i \mid c_{i-4}, \ldots, c_{i-1}) + \lambda P_{\text{ext}}(c_i \mid Z_i),
    \)
    where \( \lambda = 0.8 \) in this example. The resulting distribution could be:
    \(
    \{ \texttt{r}: 0.20, \texttt{d}: 0.24, \texttt{l}: 0.28, \ldots \}.
    \)
    
    \item \textbf{Sampling and update}: The next character (e.g., \texttt{l}) is sampled from the fused distribution and appended to the current prefix, updating it from \texttt{love} to \texttt{lovel}. This updated prefix is then used as the query input for the next iteration. To preserve knowledge diversity, the prefix is truncated to retain only the most recent 4 characters (i.e., \texttt{ovel}), ensuring the Retriever does not focus on overly specific contexts.
    
    \item \textbf{Termination}: The above steps are repeated until a complete password is generated. Generation stops when a minimum length (e.g., 5 characters) is reached or a termination symbol (e.g., \textbackslash n) is produced.
\end{enumerate}

\section{Theoretical analysis}
\label{appendix3}

Here, we present a theoretical discussion, which provides a comprehensive understanding of the theoretical advantages of KAPG.

\subsection{Information capacity}
KAPG enhances password guessing by expanding its effective information capacity to incorporate external knowledge. We analyse this limitation in terms of conditional entropy, which quantifies the uncertainty in predicting the next character \(c_i \) in the case of a prefix of \(c_{i-4}, \ldots, c_{i-1} \). 
For a traditional password guessing model (e.g., Markov), the conditional entropy of the next character \( c_i \) is: 
\(
H(c_i | c_{i-4}, \ldots, c_{i-1}, \theta), 
\) 
where \( \theta \) denotes the parameters of the model. In the KAPG's Generator (i.e., a 4th order Markov model), \( \theta \) corresponds to the transition probabilities stored in the transition matrix \( T \). This entropy reflects the model's uncertainty in predicting \( c_i \), which is limited by the information encoded in \( \theta \). For example, while \( \theta \) captures local dependencies in a leaky dataset, it may not be able to account for external influences, leading to higher entropy and lower prediction accuracy.

KAPG addresses this problem by integrating an external knowledge base \(Z \). The retrieval mechanism selects the subset of knowledge items that are most relevant to the prefix \(c_{i-4}, \ldots, c_{i-1}\), thus effectively making conditional predictions on the prefix and \(Z_i\). The information gain from this external knowledge can be quantified in terms of mutual information: 
\(
I(c_i; Z_i | c_{i-4}, \ldots, c_{i-1}, \theta)=H(c_i|c_{i-4}, \ldots, c_{i-1}, \theta)-H(c_i|c_{i-4}, \ldots, c_{i-1}, Z_i, \theta),
\)
where \( I(c_i; Z_i | c_{i-4}, \ldots, c_{i-1}, \theta) \) measures the reduction in uncertainty about \( c_i \) due to \( Z_i \), given the prefix and model parameters. In password guessing, this reduction in uncertainty is substantial when \( Z_i \) includes semantically related terms, allowing KAPG to prioritise possible characters and reduce entropy. 
Thus, the retrieval enhancement design reduces the conditional entropy \( H(c_i | c_{i-4}, \ldots, c_{i-1}, Z_i, \theta) \) by utilising external knowledge, thus expanding the effective information capacity of the model. This is achieved without modifying the internal parameters \(\theta \), as the Fusionizer module dynamically integrates \(Z_i \) through weighted probabilities. The result is a more adaptable model that captures a wider range of password patterns.

\subsection{Generalization error}

Generalization error quantifies a model's performance on unseen data, particularly when password distributions evolve over time. Statistical learning theory provides a bound that relates a model's expected error on test data to its training error, the divergence between training and test distributions, and its complexity:
\(
\mathbb{E}_{(x,y) \sim P_{\text{test}}} [L(h(x), y)] \leq \mathbb{E}_{(x,y) \sim P_{\text{train}}} [L(h(x), y)] + 
D_{\text{KL}}(P_{\text{test}} \| P_{\text{train}}) + \mathcal{C}(h),
\)
where: \( (x, y) \) represents an input-output pair (e.g., a password prefix and the next character); \( P_{\text{test}} \) and \( P_{\text{train}} \) are the probability distributions of the test and training data, respectively; \( \mathbb{E}_{(x,y) \sim P} [\cdot] \) denotes the expectation over the distribution \( P \); \( L(h(x), y) \) is the loss function, measuring the difference between the model's prediction \( h(x) \) and the true value \( y \); \( h \) is the model's hypothesis function (i.e., the mapping from input to output); \( D_{\text{KL}}(P_{\text{test}} \| P_{\text{train}}) \) is the Kullback-Leibler (KL) divergence, measuring the difference between the test and training distributions; \( \mathcal{C}(h) \) represents the model complexity term, related to capacity and regularization. 
For traditional models, when the test distribution \( P_{\text{test}} \) includes new trends not present in the training distribution \( P_{\text{train}} \), the KL divergence \( D_{\text{KL}}(P_{\text{test}} \| P_{\text{train}}) \) increases significantly, leading to a decline in generalization performance. This explains why traditional models require frequent retraining to remain effective.

KAPG mitigates this issue by adjusting its effective training distribution through the external knowledge base, resulting in the following generalization error bound:
\(
\mathbb{E}_{(x,y) \sim P_{\text{test}}} [L(h_{\text{KAPG}}(x), y)] \leq \mathbb{E}_{(x,y) \sim P_{\text{dynamic}}} [L(h_{\text{KAPG}}(x), y)] + D_{\text{KL}}(P_{\text{test}} \| P_{\text{dynamic}}) + \mathcal{C}(h_{\text{KAPG}}),
\)
where \( P_{\text{dynamic}} \) is a dynamic distribution that combines static training data with the external knowledge base. By periodically updating the knowledge base, KAPG ensures that \( P_{\text{dynamic}} \) remains close to \( P_{\text{test}} \), thereby maintaining a small KL divergence. This adjustment mechanism significantly enhances KAPG's generalization ability under evolving password distributions, reducing the reliance on retraining.

\subsection{Advantages of knowledge integration over retraining}

The knowledge integration strategy of KAPG offers significant computational and scalability advantages over retraining-based approaches in adapting to new password patterns. The traditional model integrates the new data \(D'\) by recomputing all parameters \( \theta \): 
\( 
\theta_{\text{new}} = \arg \max_{\theta} P(D \cup D' \mid \theta), 
\) 
where: \( D \) is the original training dataset; \( D' \) is the new dataset; \( D \cup D' \) is the combined dataset. The computational complexity of this retraining process is 
\(
O(|\theta| \times |D \cup D'|), 
\) 
where \( |\theta| \) is the number of parameters, and \( |D \cup D'| \) is the size of the merged data set. As the password dataset grows, this linear complexity becomes computationally expensive, especially when frequent updates are required to capture new patterns.

In contrast, KAPG uses a hierarchical knowledge representation: stable patterns are stored in the Generator's parameters, while external patterns, such as popular terms (e.g., \texttt{covid-19}), are kept in an external knowledge base. When new data \(D' \) is available, only \(Z \) is updated: 
\(
Z_{\text{new}} = \text{Update}(Z, D'), 
\) 
with a computational complexity of 
\( 
O(|D'| \times fz(|Z|)), 
\) 
where \( |D'| \) is the size of the new data; \( fz(|Z|) \) is a function relating the knowledge base size, which is usually sublinear (e.g., \( \log(|Z|) \) in the Faiss index). This efficient update process allows KAPG to incorporate new patterns without retraining the entire model, thus significantly reducing computational overhead.

\subsection{Why a simple generator suffices?}
The choice of a simple generator in KAPG that predicts the next character \( c_i \) based on the prefix \( c_{i-4}, \ldots, c_{i-1} \) is theoretically sound in terms of stability, and complementarity with retrieval. First, its simplicity ensures a stable baseline probability \( P_{\text{int}}(c_i \mid c_{i-4}, \ldots, c_{i-1})\) that captures local feature dependencies with low variance and avoids the risk of overfitting that comes with complex models. Second, it delegates the modelling of external patterns to a retriever, which uses an external knowledge base \(Z\) to incorporate influences such as popular vocabulary, reducing the complexity of the generator while improving the efficiency of the system. Third, the Fusionizer module refines this baseline by integrating retrieved knowledge \(Z_i\), iteratively improving predictions as prefixes evolve. As mentioned above, the capacity of the generator is initially limited by its parameters \(\theta\), with the number of parameters \(|\theta|\) reflecting the model's ability to encode patterns in the training data. This is approximated as number in \(T\), capacity \( \log(|\theta|) \) bits. By adding \(Z\), KAPG enhances this capacity. A simplified approximation of the extended capacity is \( \log(|\theta|) + \log(|Z|) \), where \( \log(|Z|)\) represents the potential contribution of the knowledge base. Thus, the simple generator provides a stable basis for strategically utilising retrieval enhancements for comprehensive password cracking.

\begin{figure*}[htbp]
\centering
\begin{subfigure}{0.24\linewidth}
    \centering
    \includegraphics[width=1\linewidth]{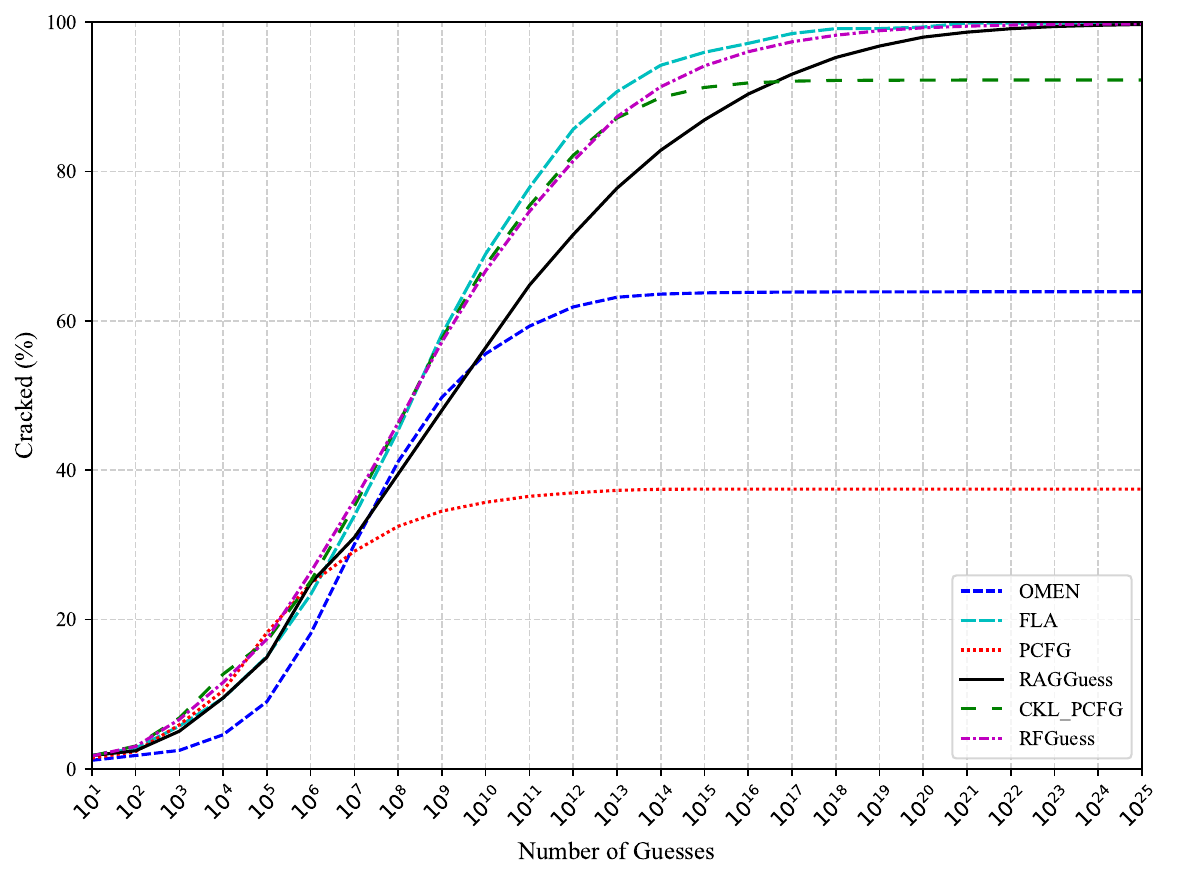}
    \caption{\textmd{0.18M Linkedin $\rightarrow$ Gmail}}
    \label{CSDN_length}
\end{subfigure}
\centering
\begin{subfigure}{0.24\linewidth}
    \centering
    \includegraphics[width=1\linewidth]{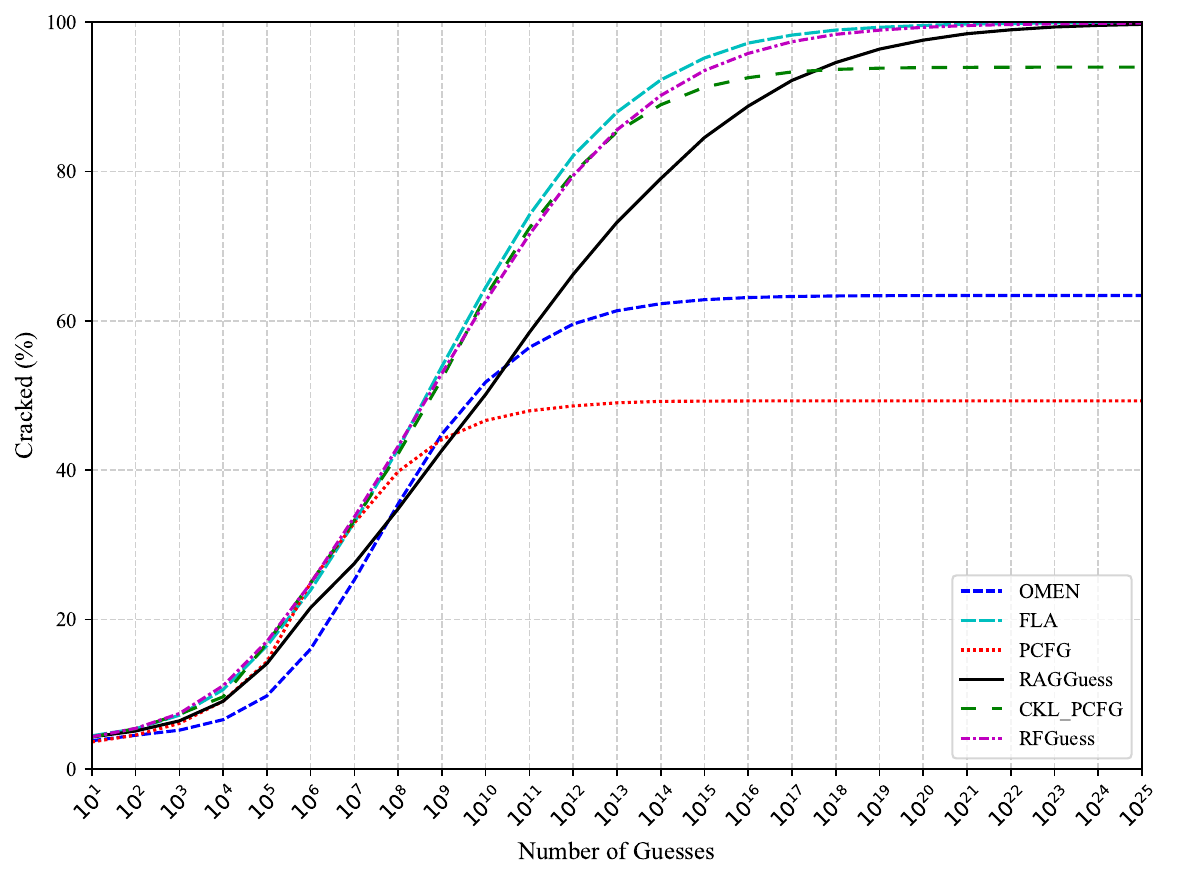}
    \caption{\textmd{0.18M Mathway $\rightarrow$ Mathway}}
    \label{CSDN_length}
\end{subfigure}
\centering
\begin{subfigure}{0.24\linewidth}
    \centering
    \includegraphics[width=1\linewidth]{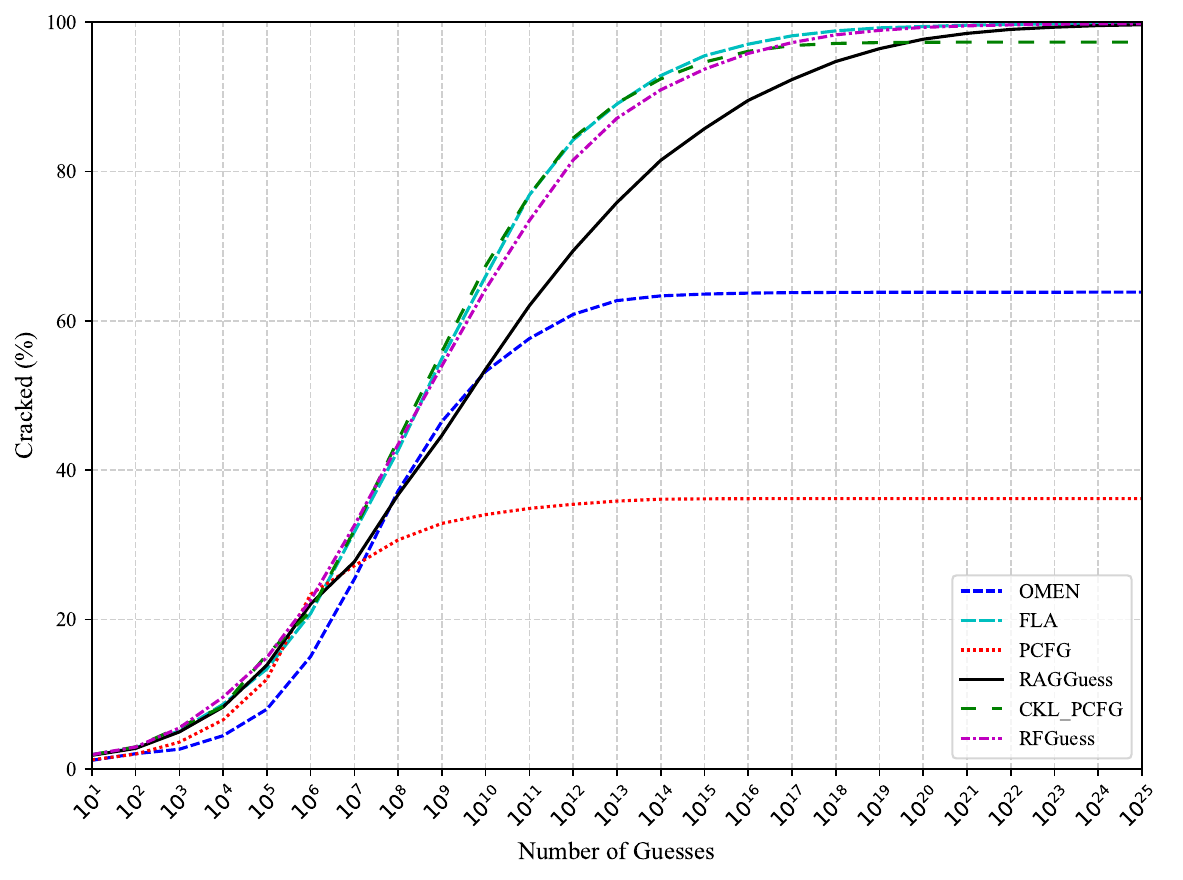}
    \caption{\textmd{0.18M Mathway $\rightarrow$ Gmail}}
    \label{CSDN_length}
\end{subfigure}
 \centering
\begin{subfigure}{0.24\linewidth}
    \centering
    \includegraphics[width=1\linewidth]{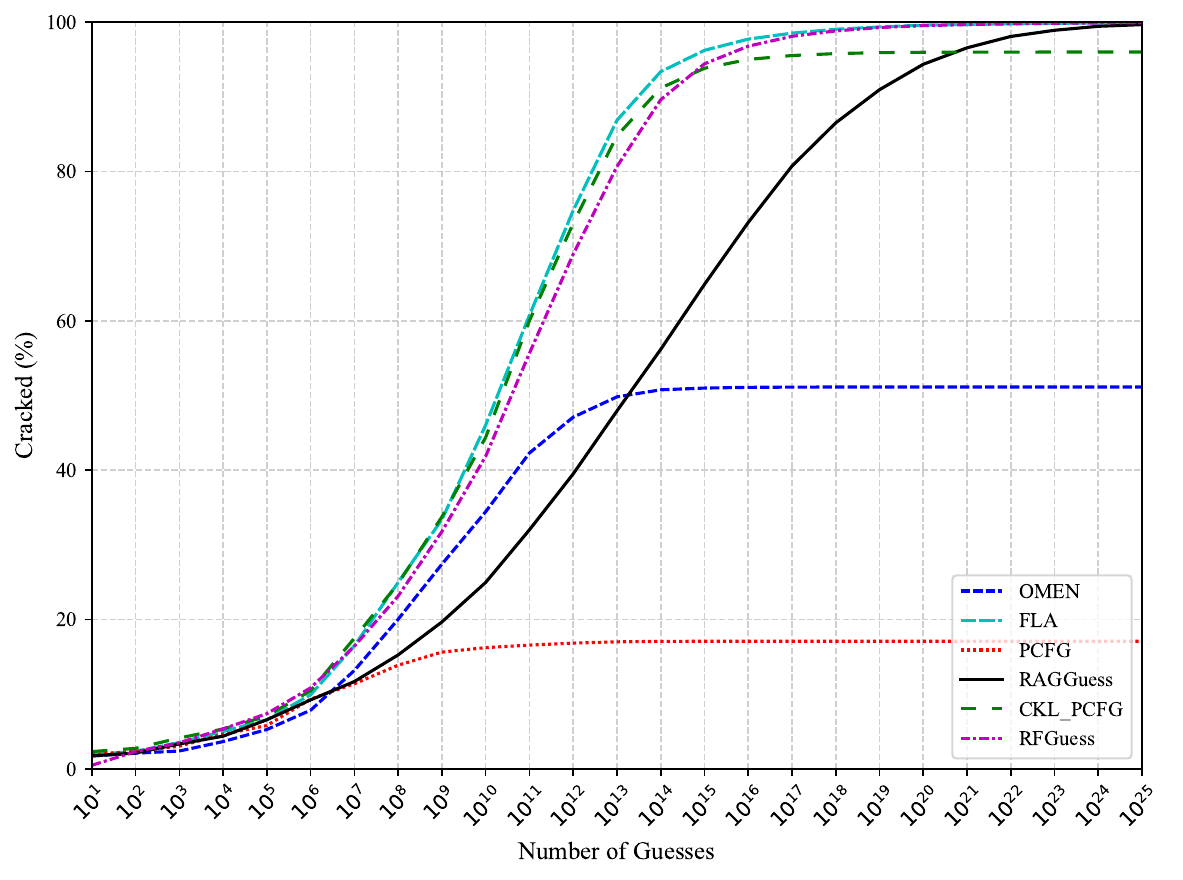}
    \caption{\textmd{0.18M Neopets $\rightarrow$ Dodonew}}
    \label{CSDN_length}
\end{subfigure}
\centering
\begin{subfigure}{0.24\linewidth}
    \centering
    \includegraphics[width=1\linewidth]{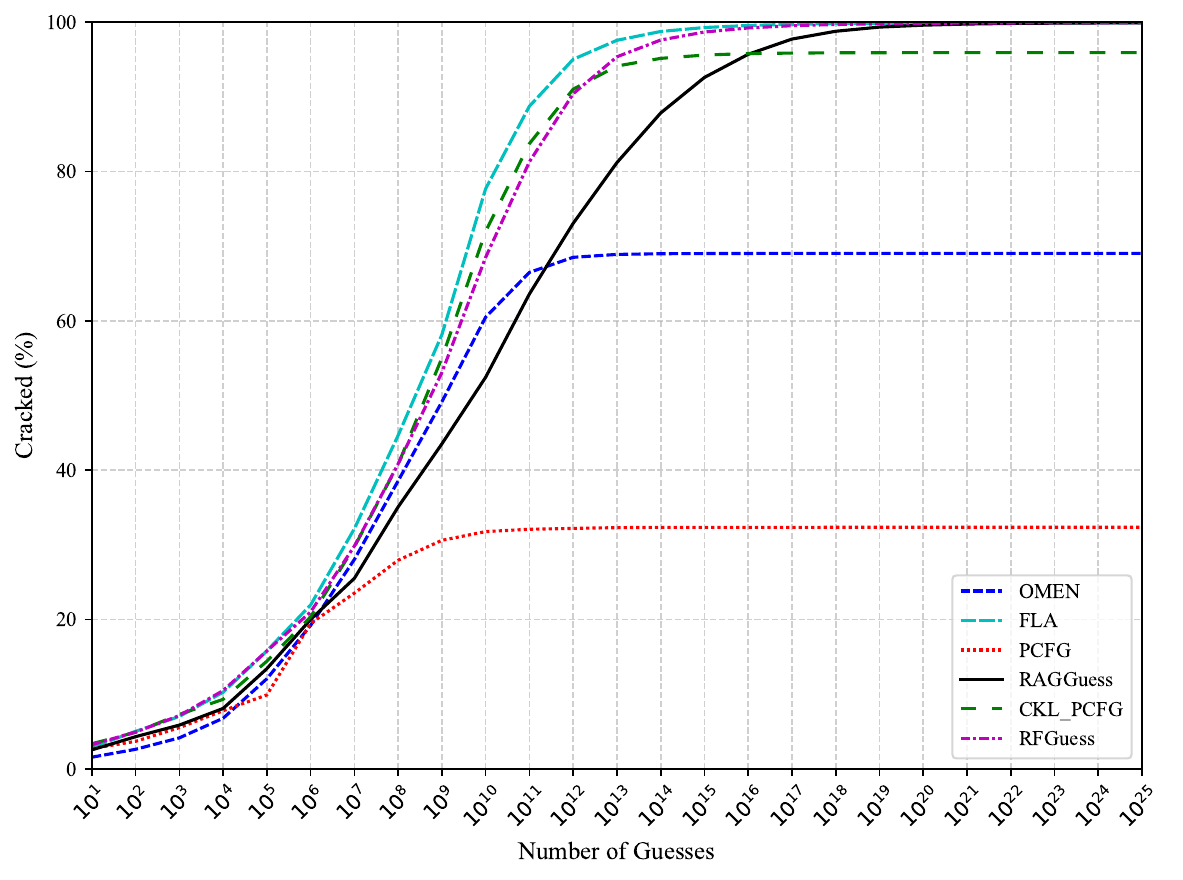}
    \caption{\textmd{0.18M Netease $\rightarrow$ Dodonew}}
    \label{CSDN_length}
\end{subfigure}
\centering
\begin{subfigure}{0.24\linewidth}
    \centering
    \includegraphics[width=1\linewidth]{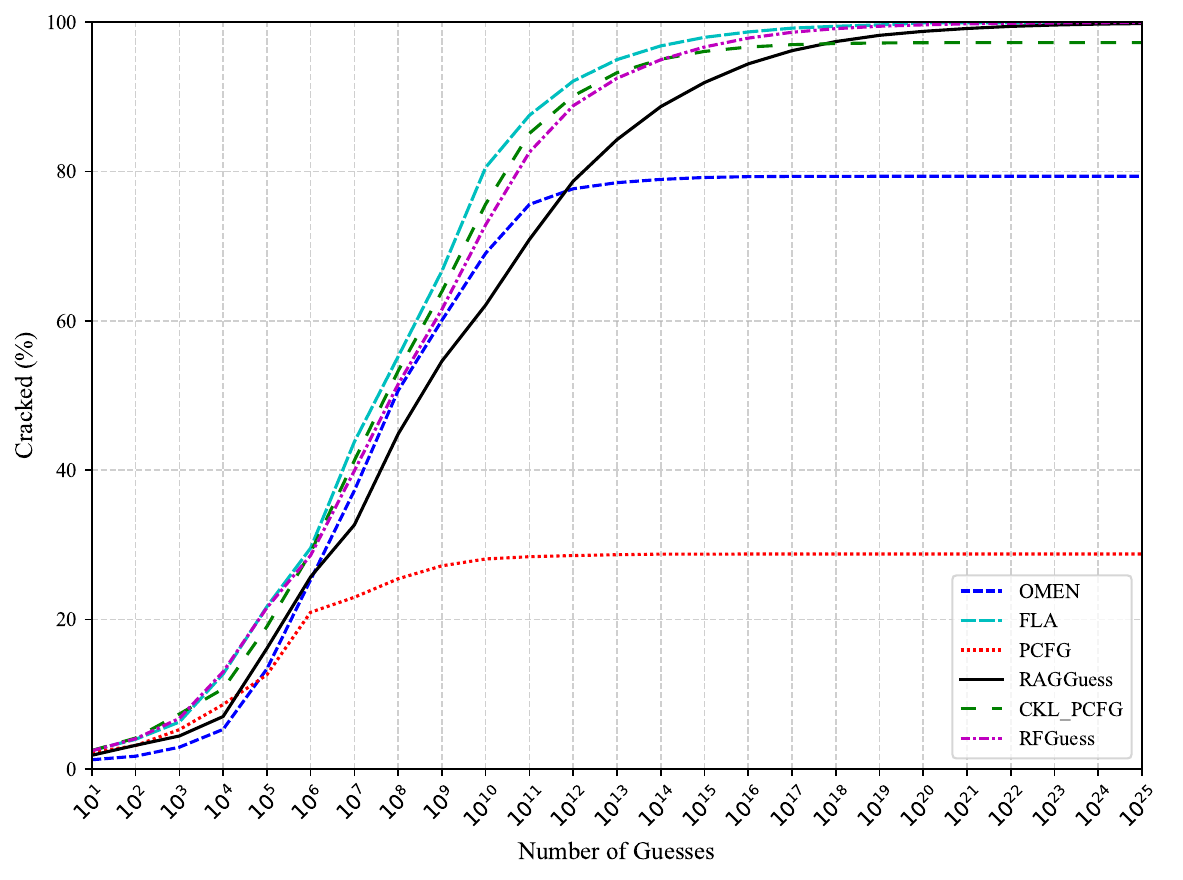}
    \caption{\textmd{0.18M Netease $\rightarrow$ Netease}}
    \label{CSDN_length}
\end{subfigure}
\centering
\begin{subfigure}{0.24\linewidth}
    \centering
    \includegraphics[width=1\linewidth]{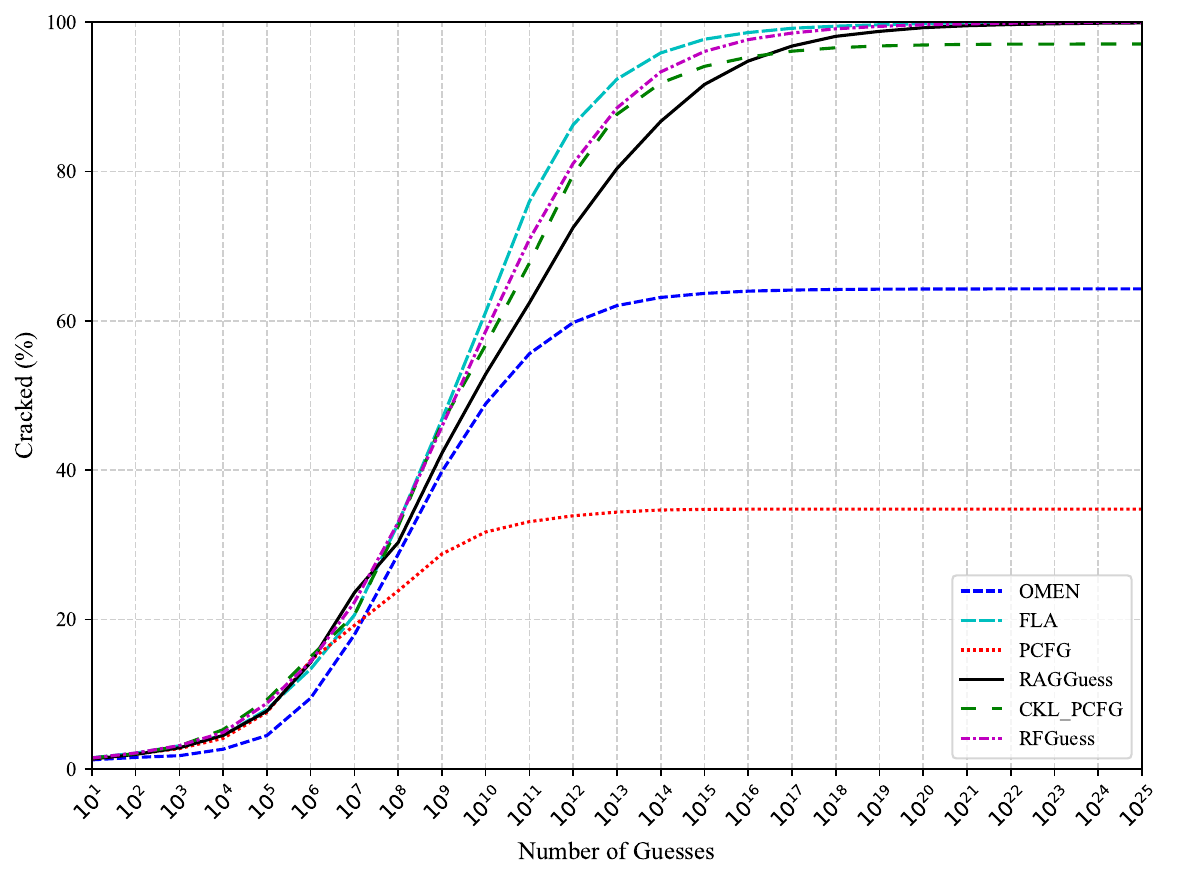}
    \caption{\textmd{1.8M 7K7K $\rightarrow$ Linkedin}}
    \label{CSDN_length}
\end{subfigure}
\centering
\begin{subfigure}{0.24\linewidth}
    \centering
    \includegraphics[width=1\linewidth]{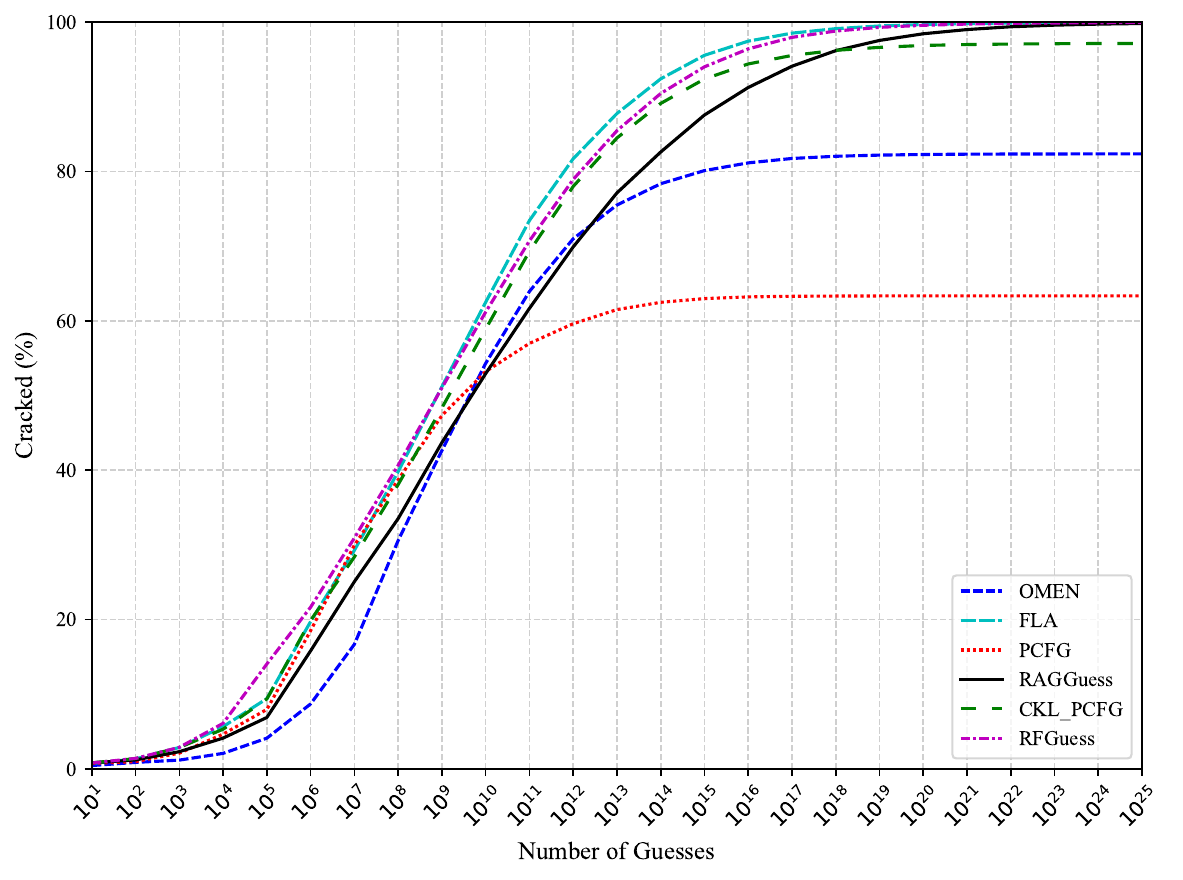}
    \caption{\textmd{1.8M Clixsense $\rightarrow$ Mathway}}
    \label{CSDN_length}
\end{subfigure}
\centering
	\begin{subfigure}{0.24\linewidth}
		\centering
		\includegraphics[width=1\linewidth]{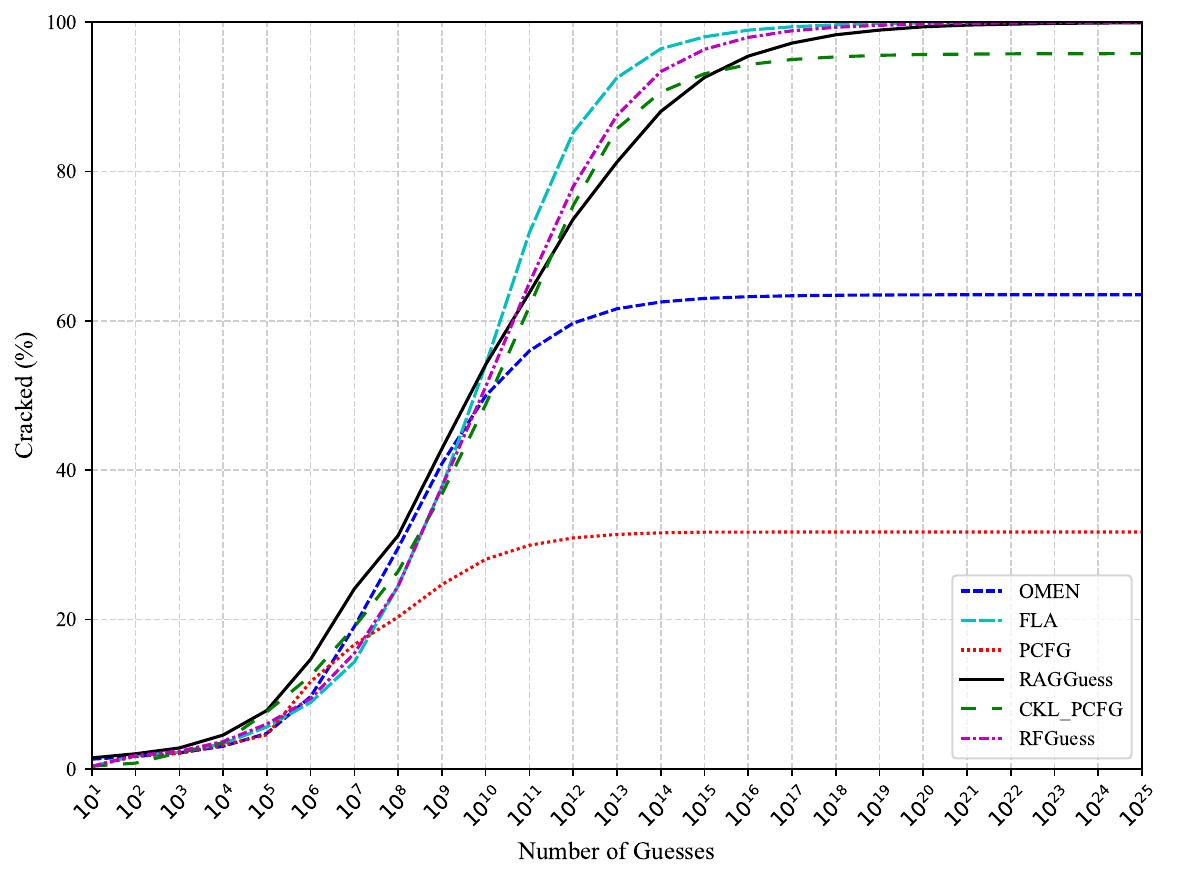}
		\caption{\textmd{1.8M CSDN $\rightarrow$ Linkedin}}
		\label{CSDN_length}
	\end{subfigure}
\centering
	\begin{subfigure}{0.24\linewidth}
		\centering
		\includegraphics[width=1\linewidth]{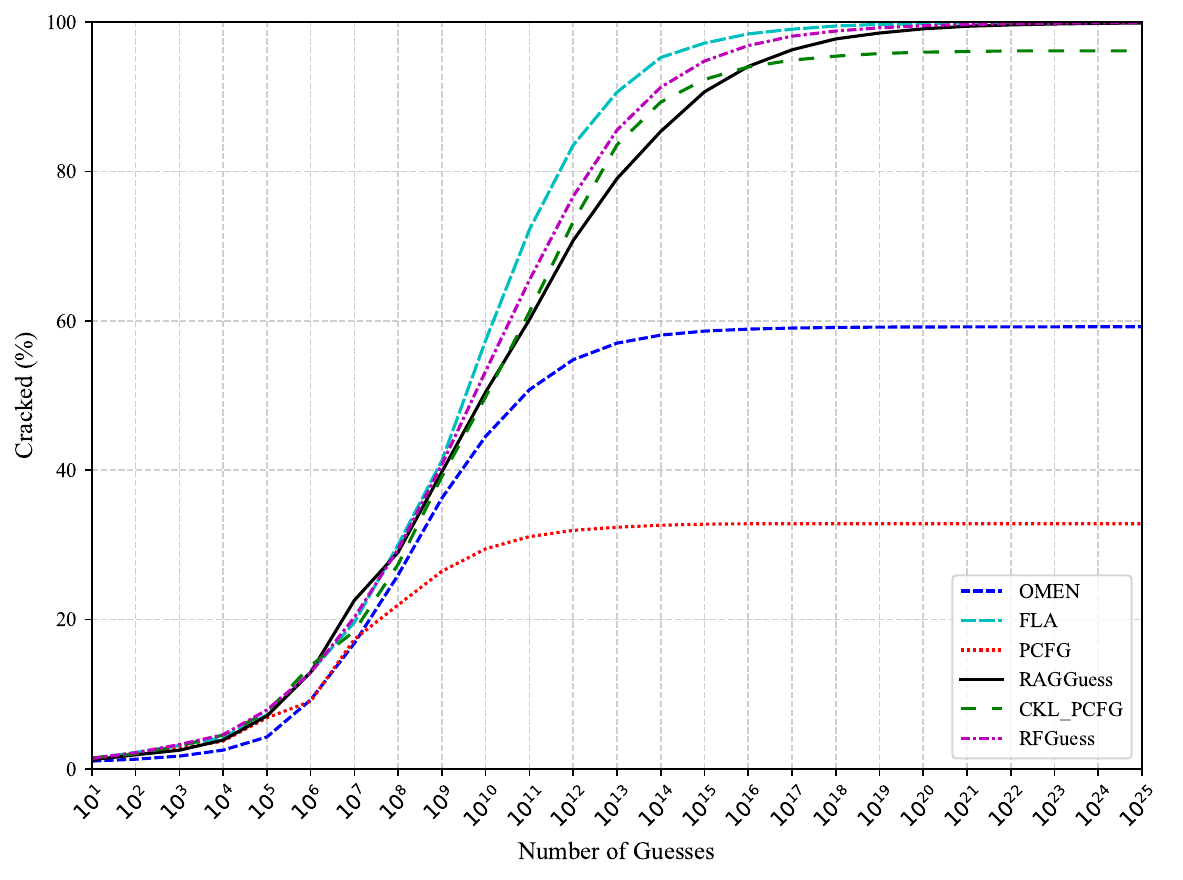}
		\caption{\textmd{1.8M Netease $\rightarrow$ Linkedin}}
		\label{CSDN_length}
	\end{subfigure}
        \centering
	\begin{subfigure}{0.24\linewidth}
		\centering
		\includegraphics[width=1\linewidth]{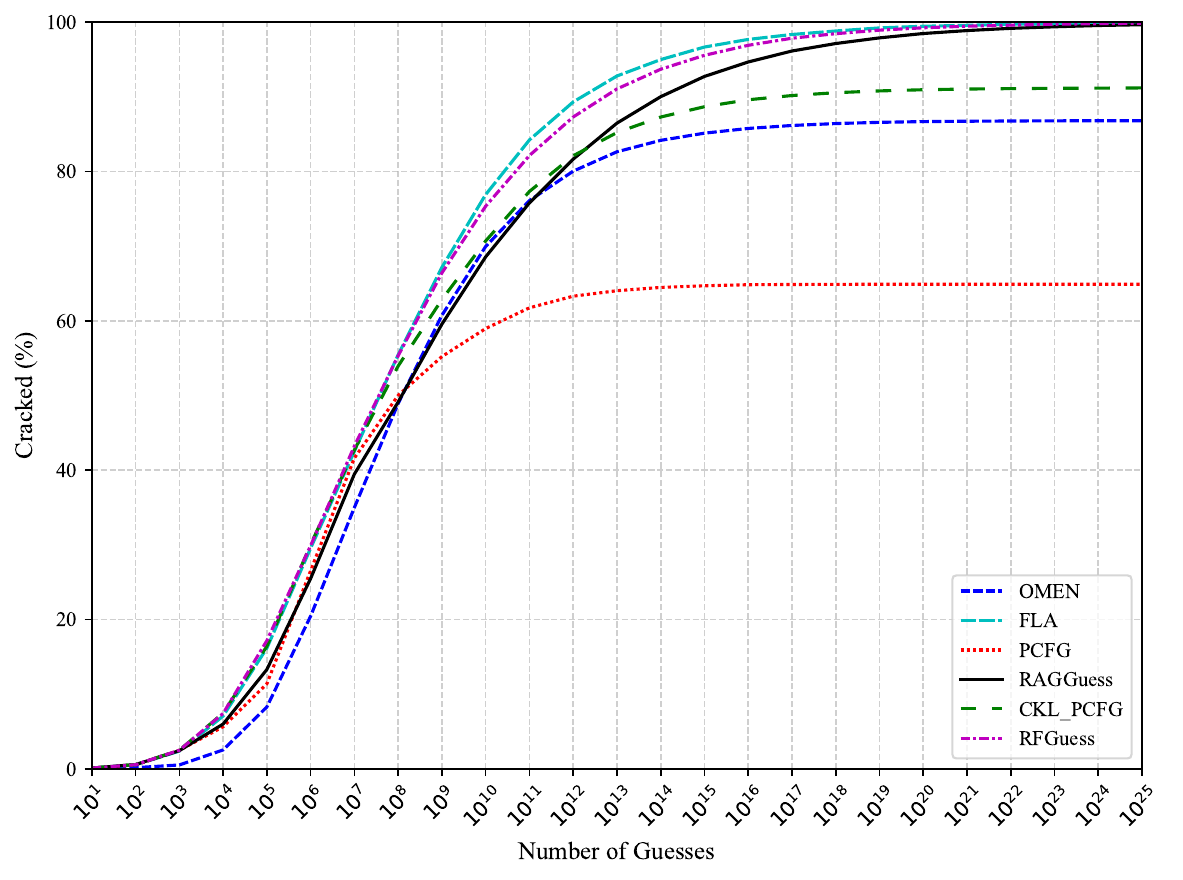}
		\caption{\textmd{1.8M Rockyou $\rightarrow$ Neopets}}
		\label{CSDN_length}
	\end{subfigure}
        \centering
	\begin{subfigure}{0.24\linewidth}
		\centering
		\includegraphics[width=1\linewidth]{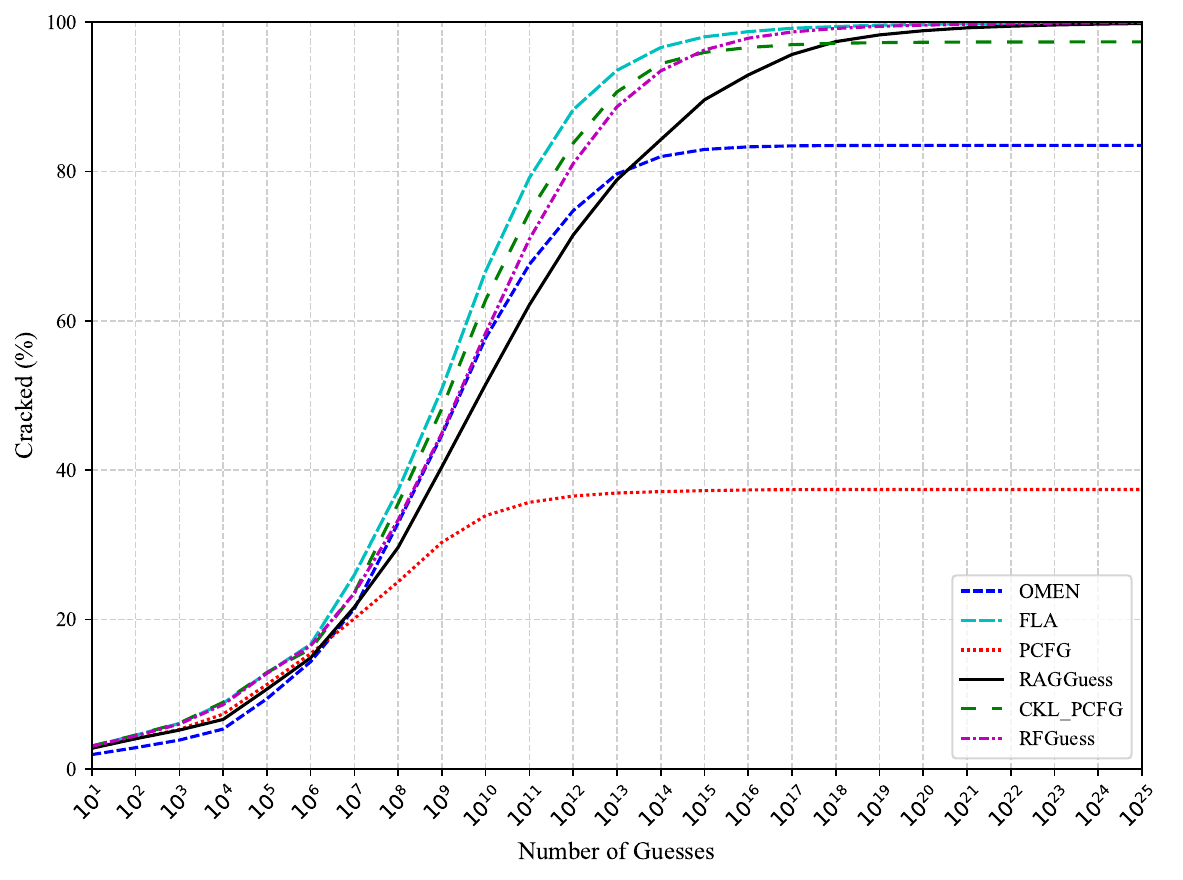}
		\caption{\textmd{1.8M Tianya $\rightarrow$ Taobao}}
		\label{CSDN_length}
	\end{subfigure}
\caption{Supplementary experimental results.}
	\label{fig:supplementary guessing result}
\end{figure*}

\begin{algorithm*}[!ht]
\caption{Password strength estimation algorithm (KAPSM)}
\label{alg:ragpsm}
\renewcommand{\algorithmicrequire}{\textbf{Input:}}
\renewcommand{\algorithmicensure}{\textbf{Output:}}
\begin{algorithmic}[1]
    \REQUIRE A password as a sequence of characters $p = c_1 c_2 \dots c_L$, starting prefix and KAPG model, including Generator, Fusionizer, and Retriever modules
    \ENSURE Estimated strength $\text{guess}_{\text{number}}$ required to crack the password
    
    \STATE Initialize $\text{total}_{\text{prob}} \gets 1$;
    \STATE $\text{prefix} \gets \text{starting prefix}$;
    \FOR{each $i \in [1,L]$}
    
    \STATE $P_{\text{int}} \gets \text{Generator.predict\_next}(\text{prefix})$; \/// predict next character distribution
    
    \STATE $Z_{i} \gets \text{Retriever.retrieve\_topk}(\text{prefix})$; \/// retrieve top-10 external knowledge items
    \STATE $P_{\text{ext}} \gets \text{aggregate\_distribution}(Z_{i})$; \/// aggregate external distribution
    
    \STATE $\lambda_i \gets \text{Fusionizer}(\text{sim}(Z, \text{prefix}))$; \/// compute fusion weight
    \STATE $P_{\text{fused}} \gets (1 - \lambda_i) \cdot P_{\text{int}} + \lambda_i \cdot P_{\text{ext}}$; \/// fused probability distribution
    
    \STATE $\text{total}_{\text{prob}} \gets \text{total}_{\text{prob}} \cdot P_{\text{fused}}[c_i]$; \/// update with probability of current character
    \STATE $\text{prefix} \gets \text{prefix}+c_i$;
    \ENDFOR
    
    \STATE $\text{guess}_{\text{number}} \gets \text{estimate\_guesses}(\text{total}_{\text{prob}})$; \/// convert probability to guess count via monte carlo
    \RETURN $\text{guess}_{\text{number}}$;
  \end{algorithmic}
\end{algorithm*}

\end{document}